\documentclass[a4paper,11pt]{article}
\usepackage[utf8]{inputenc}
\usepackage[T1]{fontenc}
\usepackage[english]{babel}
\usepackage{geometry} 
\usepackage{marginnote}
\usepackage{mathtools}
\usepackage{graphicx}
\usepackage{amsmath}
\usepackage{amsfonts}
\usepackage{amssymb}
\usepackage{pstricks-add}
\usepackage{multicol,caption}
\usepackage{geometry}
\usepackage{caption}
\usepackage{lipsum}
\usepackage{tabularx}
\usepackage{multirow}
\usepackage{placeins}
\usepackage{subfigure}
\usepackage{textcomp}
\usepackage{hyperref}
\usepackage{amsfonts}
\usepackage{amsbsy}
\usepackage{svg}
\usepackage{textgreek}
\usepackage{scalerel}
\usepackage{ifthen}

\usepackage{subfigure}

\author{
  Eric Behle$^{1}$, Adélaïde Raguin$^{1}$
  \\[0.5em]
  \begin{minipage}{0.9\textwidth}\small
  $^1$ Department of Biology, Cluster of Excellence on Plant Sciences, Institute of Quantitative and Theoretical Biology, Heinrich-Heine University, 40225 Düsseldorf, Germany
\\\\
\normalsize Corresponding author: adelaide.raguin@hhu.de
\end{minipage}
}

\title{Stochastic Model of Lignocellulosic Material Saccharification}

\begin{document}
\maketitle

\newenvironment{Figure}
{\par\medskip\noindent\minipage{\linewidth}}
{\endminipage\par\medskip}

\abstract{
\noindent The processing of agricultural wastes towards extraction of renewable resources is recently being considered as a promising alternative to conventional biofuel production processes. Agricultural residues represent an abundant and unexploited raw material that intrinsically contains chemical energy in the form of polysaccharides. The degradation procedure is a complex chemical process that is currently time intensive and costly. Various pre-treatment methods are being investigated to determine the subsequent modification of the material and the main obstacles in increasing the enzymatic saccharification yield. In this study, we present a computational model that complements the experimental approaches. We decipher how the three-dimensional structure of the substrate impacts the saccharification yield. We model a cell wall microfibril composed of cellulose and surrounded by hemicellulose and lignin, with various relative abundances and arrangements. This substrate is subjected to digestion by different enzyme cocktails of well characterized enzymes. The saccharification dynamics is mimicked \textit{in silico} using a stochastic simulation procedure based on a Gillespie algorithm. As we additionally implement a fitting procedure that optimizes the parameters of the simulation runs, we are able to reproduce experimental saccharification time courses for corn stover. Our results challenge the hypothesis of hemicellulose content being a substantial factor in saccharification yield, and instead propose the crystallinity of the substrate having a much higher impact.

\begin{center}
\textbf{Author summary}
\end{center}
Leftover wastes generated by agriculture, such as inedible leaves and stalks of plants, represent an abundant and unexploited raw material that contains energy in the form of sugar polymers. Their breakdown and processing into bio-ethanol is recently being considered as a promising candidate for renewable fuel production. However, it is still poorly understood, how the microscopic structure and composition of plant waste materials impact their enzymatic digestion. Various experimental pre-processing methods are currently being tested to determine their effect on the material composition and structure, and the sugar yield. In this study, we present a computational model to complement such experimental approaches. We simulate a microscopic plant fragment typically found in plant waste materials, whose structure and composition can be tailored. This fragment is then subjected to enzymatic digestion, whose dynamics is tracked in silico. The model reproduces experimentally observed time courses for plant fragments of known composition. It additionally provides new hypotheses for interpreting complex experimental results.

}

\begin{multicols}{2}

\section{Introduction}

\noindent The worldwide challenges of energy supply and resource shortage are becoming ever more urgent. Fully utilizing renewable resources such as waste biomaterials generated via the agricultural industry is an inviting alternative to face these difficulties \cite{Carroll2009}. Lignocellulosic materials which are currently considered waste, for example those parts of crops which are not applicable for animal or human consumption, are of particular interest. They contain large amounts of chemical energy, which may be utilized in processing sectors such as the biofuel industry. Strategies for energy extraction include high-temperature conversion to syngas, fast pyrolysis, and bioconversion methods \cite{Carroll2009}, each of which is still an area of active research. We focus here on the bioconversion methods. The objective is to extract monomeric sugars from the many polysaccharides found within lignocellulosic material, and then to ferment them into ethanol. The process of extracting these sugars is called saccharification, and is carried out via enzymatic digestion.

\noindent Efficient enzymes for sugar extraction can be found in microorganisms which are specialized towards plant digestion \cite{Beldman1987}. Utilizing them for human benefit has been studied extensively and these enzymes are well characterized \cite{Beldman1987, Divne1994}. However, this strategy is so far costly, as it requires isolation of large quantities of the enzymes involved \cite{Carroll2009}. As a result, various pre-treatment methods have been conceived \cite{Weidener2020, Bura2009, Kucharska2018}. Their purpose is to alter the substrate structure in order to improve ease of access for the enzymes to the valuable sugars. The methods used include pre-treatment with acids such as H$_2$SO$_4$ or HCl, as well as exposure to steam at various temperatures and exposure times. The development of these methods in turn has led to questions regarding the effect of the substrate structure on the action of the enzymes, as well as their interaction with non-digestible lignocellulose components.\\

\noindent Experiments have measured the saccharification dynamics of lignocellulosic material for different pre-treatment severities \cite{Bura2009}, and determined the impact of structural properties of the substrate (particle size \cite{Zoghlami2019}, crystallinity \cite{Weidener2020, Pena2019, Thomas2013}). To complement this, modeling approaches have been used to simulate the static structure of lignocellulosic materials. Charlier and Mazeau \cite{Charlier2012} proposed a molecular dynamics model of the cell wall components to investigate their arrangement depending on their chemical properties. Yi and Puri \cite{Yi2012} investigated the mechanical properties of the cell wall by using the finite element method, and Ciesielski et al.\cite{Ciesielski2020} combined structure modeling approaches at different length scales. Additionally, a few studies also address the saccharification dynamics. Vetharaniam et al. \cite{Vetharaniam2014} developed a 3-D, agent-based model of cell wall digestion, and Kumar and Murthy \cite{Kumar2017} combined experimental and theoretical (Monte Carlo) methods to investigate the effect of enzyme crowding.\\

\noindent Following a purely theoretical approach supplemented by the abundant experimental data found in the literature, in this study we present a comprehensive computational model based on stochastic simulations. It is a coarse-grained and simplistic representation of the system whose purpose is to capture essential biological features and to analyze them in depth by comparison to experimental data. Our aim is to understand the effect of the substrate structure on the action of the enzymes as well as their interaction with non-digestible lignocellulose components. Thus, the model accounts both for the composition and three-dimensional structure of the substrate, and the distinct enzymes typically used as cocktails in industry. The model contributes to decipher the interplay between the enzymes and the substrate during the saccharification process. This means not only to track how the enzymes synergistically digest the substrate over time, but also to characterize the impact of the substrate and its spatial structure on the action of the enzymes. The model explains the synergism of the enzymes involved for a fast and complete saccharification. It shows a linear decrease of the saccharification yield with the lignin content as well as with the crystallinity of the substrate. These two qualitative tendencies are supported by experimental evidence. Additionally, we semi-quantitatively validate the model by comparing our simulations to experimental saccharification curves for distinct pre-treatments of corn stover samples. As we reproduce these experimental data, we demonstrate the importance of the substrate structure and crystallinity to elucidate the results. Our theoretical findings also question the role of the substrate composition, so far considered as predominant in the interpretation of the experimental data fitted here.\\

\noindent The manuscript is organized as follows: in section \ref{The_model} we present the model in detail in terms of the underpinning biological system and its computational implementation. This includes to briefly review the principles of the stochastic Gillespie algorithm. In section \ref{Model_analysis}, we present the model's results in light of experimental measurements. We decipher the enzyme activity and characterize the influence of the lignin content and the crystallinity of the substrate on the saccharification process and yield. Then, we utilize published experimental data by Bura et al. \cite{Bura2009} to rationalize saccharification time courses for corn stover materials pre-treated to different extent. This enables us to infer the critical role of substrate structure and, in particular, substrate crystallinity. In section \ref{discussion_conclusion}, we discuss our results and consider further potential developments of our modeling approach. \\ 

\noindent Extensive technical details on our computational approach are provided on our public GitLab repository ("\url{https://gitlab.com/erbeh/pcwsm}"). This includes details on the Gillespie algorithm, clean and commented codes, instructions on how to run them, explanations of the content of the input and output files of the simulation code, and details on our optimization procedure to fit experimental saccharification curves. Also, we provide all \textit{in silico} data and scripts necessary to produce the figures presented here.

\section{The model}\label{The_model}

\noindent The extraction of simple sugars from complex polysaccharides results from the combined action of enzymes on the substrate. Plant cell wall microfibrils are an insoluble and structured three-dimensional cellulosic material whose composition and arrangement together with that of hemicellulose and lignin impact the enzymatic process \cite{Weidener2020}. In light of the challenges for the saccharification of plant waste material, our aim in this study is to unravel the impact of the substrate structure on the saccharification process. To do so, we build a theoretical model that complements experimental approaches. In the model, individual parameters of the system can be varied and measured in an independent and fully controlled manner. Thus, the model also constitutes a general platform to investigate different plant mutants, tissues and enzyme abundances and kinetics. In this section, we provide details on the biological system and features included in the model in terms of substrate and enzymes. We also clarify the assumptions we make towards reducing the computational cost of the model while retaining the features we consider essential for the investigation of the impact of the substrate structure on the saccharification process. Finally, we briefly introduce the simulation procedure, i.e., the Gillespie algorithm for stochastic simulations. \\

\subsection{Plant cell wall composition and structure}\label{Substrate structure}

\noindent The plant cell wall is a complex structure usually consisting of a primary and secondary layer \cite{Gibson2012}. Each layer contains different amounts of proteins and structural polymers. The microstructure, i.e., the composition and arrangement of the components, varies strongly between different cell types, and both the individual cell composition and the multicellular arrangement strongly influence the mechanical properties of the plant \cite{Gibson2012}. In this study we focus on the three types of biopolymers found primarily within plant cell walls: cellulose, hemicellulose and lignin. Considering an averaged plant cell wall composition is in line with the field of applications connected to our study (biofuel industry), in which the primary and secondary cell walls are not distinguished.\\

\noindent  Cellulose is the most abundant organic polymer found on earth \cite{Gibson2012}. It functions as a structural backbone within the cell wall and is composed of linear chains of up to 15\,000 glucose molecules connected via \textbeta $(1\rightarrow 4)$ glycosidic bonds \cite{Gibson2012}. Cellulose polymers adopt a linear shape even at high degree of polymerization, and therefore represent a favorable structural backbone due to their rigidity. For further stabilization of the linear chains, bundles of multiple cellulose polymers are formed within the cell wall, so-called microfibrils \cite{Gibson2012}. These are distributed throughout the cell wall with varying degrees of order in their alignment \cite{Thomas2013}.\\

\begin{figure*}
  \centering

    	\includegraphics[width=0.9\textwidth]{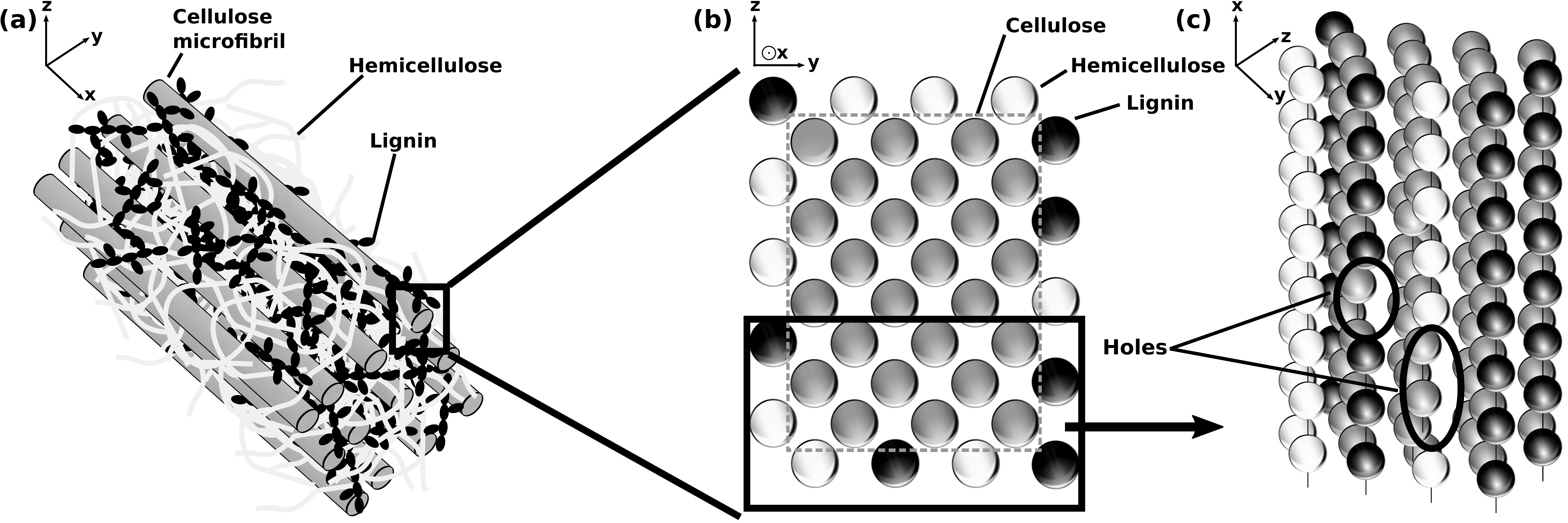}

  \captionof{figure}{
  	Schematic representation of the structure of the substrate in the model. Polymer types are color-coded (cellulose: gray; hemicellulose: white; lignin: black). (a) Several cellulose microfibrils embedded in a matrix of hemicellulose reinforced by lignin. In (b) and (c), bead chains represent monomers. (b) Top-down view of a single rectangular cellulose microfibril made of 24 polymers, and part of the surrounding matrix showing the relative arrangement of the hemicellulose and lignin polymers. (c) Side-view of the polymers highlighted in (b) by a frame, showing typical holes in the hemicellulose and lignin that surround the cellulose microfibril. 
}

  \label{Microfibril_structure}     	
\end{figure*}

\noindent  
Hemicellulose polymers consist of short amorphous chains of multiple types of sugar monomers, whose total number per polymer lies between 500 and 3\,000 \cite{Gibson2012}. Their function within the cell wall matrix is the anchoring of the cellulose microfibrils embedded within it (see Figure \ref{Microfibril_structure} (a)). The definition of hemicellulose is not fully consistent across literature. Therefore, we use that of Pauly et al. \cite{Pauly2013}, where any non-cellulose cell wall polysaccharide whose dominant backbone is linked by \textbeta $(1\rightarrow 4)$ glycosidic bonds is considered hemicellulose. Several distinct polysaccharides fit this definition. As a result, hemicellulose is further divided into four basic types \cite{Ebringerova2005}: xylans, mannans, mixed linkage \textbeta -glucans, and xyloglucans. Each of them is characterized by different sugar composition, as well as differences in their three-dimensional structure. Here we focus on a single type of hemicellulose: the xylans. Xylans are the most abundant hemicellulose polymers \cite{Prade1996}. They are characterized by a backbone composed of the five-carbon monosaccharide xylose \cite{Pauly2013}, which may be partially branched by glucoronic acid or L-arabinose, via \textalpha $(1\rightarrow 6)$ glycosidic bonds \cite{Pauly2013}. Xylans mainly occur within the secondary cell wall and act as further structure reinforcement \cite{Pauly2013}.\\

\noindent Lignin is the second most abundant biopolymer after cellulose \cite{Boerjan2003}. It is a manifold branched polymer, which is known for being a strong contributor to the mechanical properties of wood. The lignin structure is mainly composed of three monomers (monolignols) \cite{Boerjan2003}. The most common monolignols are three alcohol molecules differing only in their degree of methoxylation. \\

\noindent In the model, we retain the main features of the cellulose, hemicellulose and lignin polymers. The simulated substrate is composed of a cellulose microfibril around which hemicellulose and lignin are randomly arranged and form up to two outer layers (see Figure \ref{Microfibril_structure} (b)). The outer layers are not fully covering, but possess holes (see Figure \ref{Microfibril_structure} (c)), which represent the non-complete surrounding of the cellulose by hemicellulose and lignin. The length of the microfibril is specified in terms of the number of glycosidic bonds within an individual cellulose polymer at the start of the simulation. We set it to 100 bonds, unless otherwise specified. The shape of the microfibril, as well as the number of cellulose polymers within it, may also be specified in our model. We choose a microfibril composed of 24 cellulose polymers, arranged in a rectangular fashion, according to the most probable conformation found in plant cell walls argued by Thomas et al. \cite{Thomas2013} (see Figure \ref{Microfibril_structure} (b)).\\

\noindent The structure of the substrate is resolved in three dimensions at the scale of monomers: glucose for cellulose and xylose for hemicellulose. For lignin, monomers are a representative monolignol, which is sufficient to retain the effects of lignin we investigate (enzyme adhesion and structure blocking, see section \ref{Saccha_challenges}). For the sake of simplicity, all polymers are modeled as linear ones. The composition of the substrate in terms of polymer percentages may be tuned, such that the percentages of cellulose, hemicellulose and lignin sum up to 1. In the initial system state, the cellulose polymers have the same length, but due to the holes within the surrounding layers the hemicellulose and lignin polymers do not. All bonds within the cellulose and hemicellulose polymers are potentially subject to enzymatic digestion, unlike lignin that cannot be digested by the enzymes considered here. A bond can only be digested if it is located at an accessible position, which is the case as soon as it is exposed to the medium surrounding the microfibril (see Figure \ref{accessibility_of_bonds}).\\

\begin{Figure}
  
	\centering
    	\includegraphics[width=0.47\textwidth]{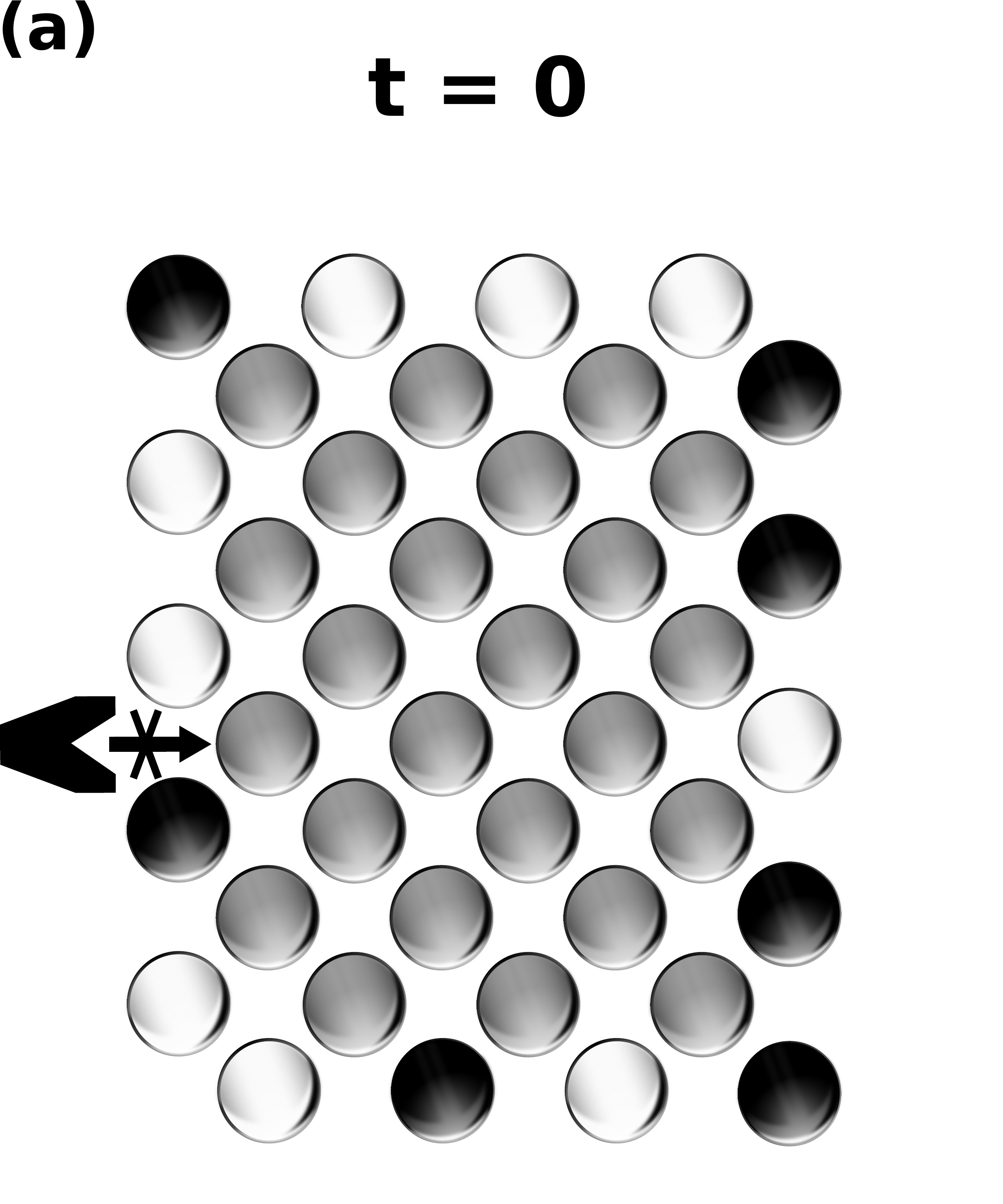}
    	\includegraphics[width=0.47\textwidth]{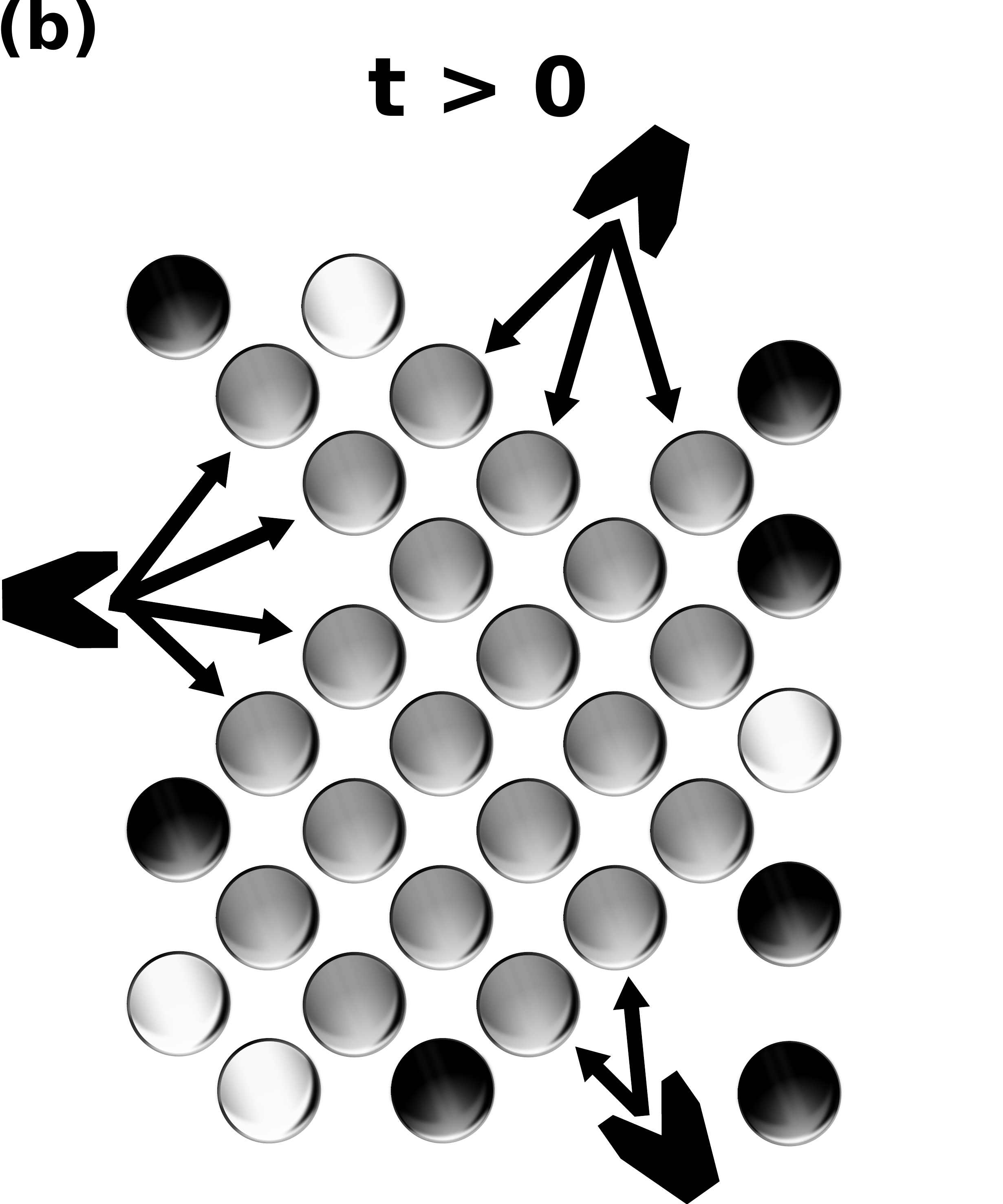}

  \captionof{figure}{
  	Sketch (top-down view) of the accessibility of polymer bonds depending on their position within the microfibril and its digestion state. (a) Beginning of the simulation: the outer layer of hemicellulose (white beads) and lignin (black beads) does not contain any holes. None of the cellulose bonds (gray beads) are accessible for digestion by enzymes (black polygons). Only the hemicellulose bonds are digestible. (b) Later stage of the simulation: some of the hemicellulose within the outer shell has been digested. The cellulose bonds highlighted by arrows are now accessible for digestion.
  }

  \label{accessibility_of_bonds}     	
\end{Figure}

\subsection{Enzyme cocktails}\label{DegradationSection}

\noindent Enzymes are a central focus for the optimization of plant cell wall saccharification. Several digesting enzymes have been isolated from nature and characterized, such that their mode of action on simple linear and soluble polysaccharides are well known. For instance, species belonging to the \textit{Trichoderma} genus excrete two groups of enzymes, respectively called cellulases and xylanases \cite{Beldman1987}. These sets of enzymes are typically used by industry to process raw plant material \cite{Straathof2014}. So, we choose to include them in the model. \\

\noindent Endoglucanases are capable of binding to any position along a cellulose polymer and digesting the \textbeta $(1\rightarrow 4)$ glycosidic bond between two neighboring glucose molecules \cite{Beldman1987, Joergensen2007}. They thereby cut the polymer into two shorter ones. The behavior of endoglucanases with respect to short polymers is not well understood. However, Scapin et al. \cite{Scapin2017} have described a bacterial endoglucanase which only digests cellulose polymers of a length of at least five glucose units. Cellobiohydrolases specifically digest the glycosidic bonds at either the reducing or non-reducing ends of cellulose polymers, and cut off cellobiose \cite{Beldman1987, Joergensen2007}. This is done in a processive manner \cite{Brady2015}. \textbeta -glucosidases complete the saccharification process by digesting cellobiose into two glucose molecules \cite{Chirico1987, Joergensen2007}. Xylan-type hemicellulose is digested by xylanases, a group of enzymes which act analogously to the cellulases on cellulose, and whose action leads to the release of single xylose molecules \cite{Prade1996}.\\

\noindent

\begin{figure*}
  \centering
   \subfigure[]{
  	\includegraphics[width=0.31\textwidth]{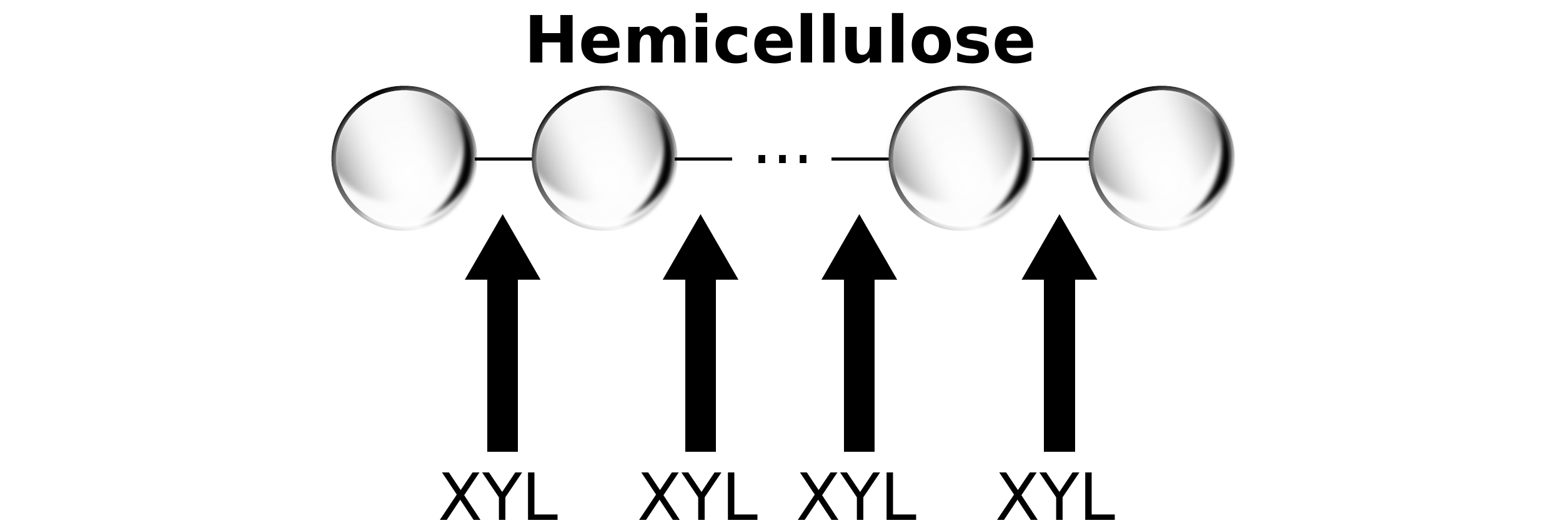}
  }
   \subfigure[]{
  	\includegraphics[width=0.31\textwidth]{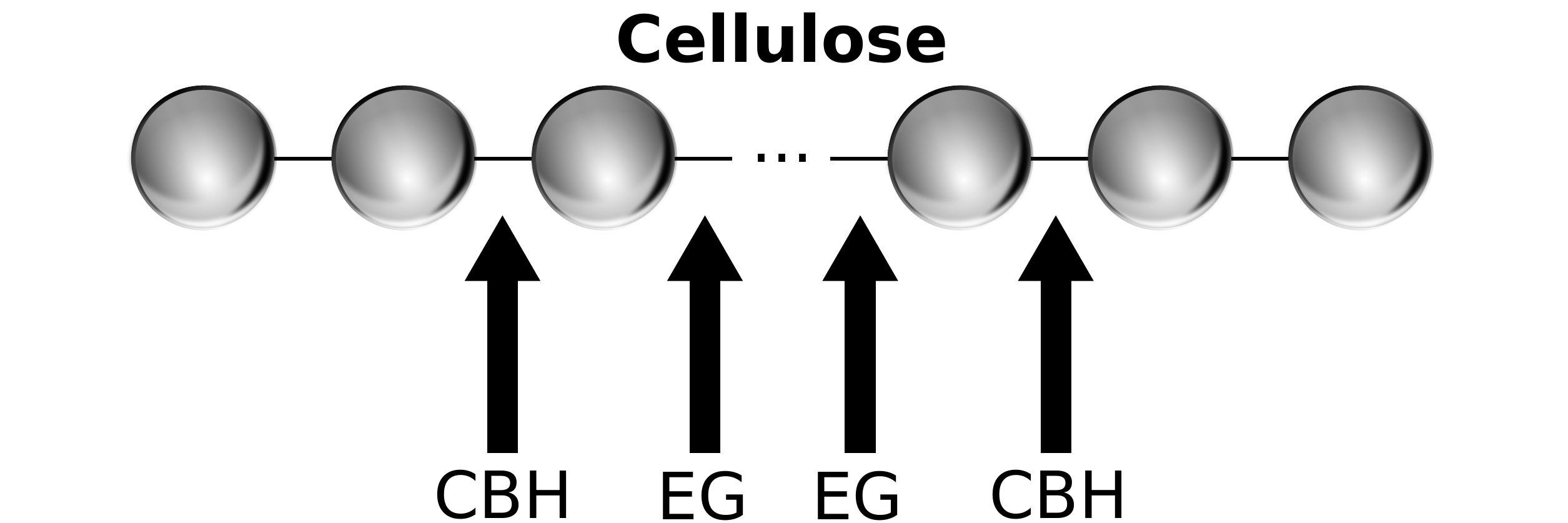}
  }
   \subfigure[]{
  	\includegraphics[width=0.31\textwidth]{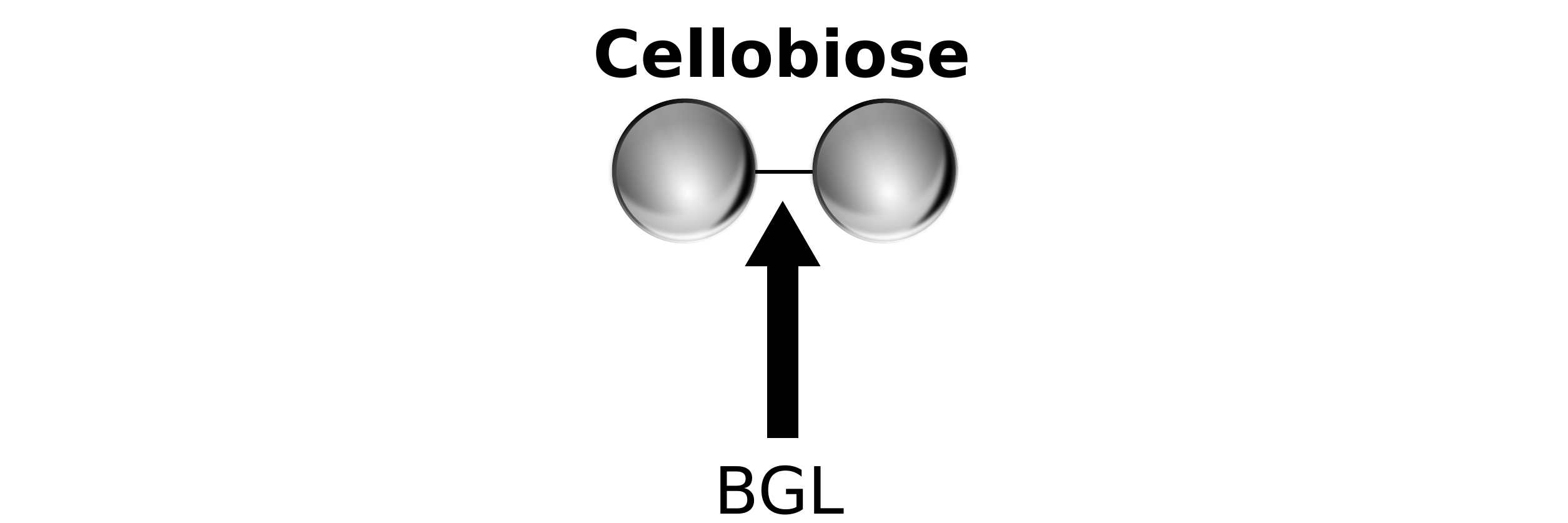}
  }

  \captionof{figure}{
Schematic representation of the polymer bonds that can be digested within the model. (a) Hemicellulose digestion sites by xylanase (XYL). Xylanase may digest xylose-xylose bonds at any position along the hemicellulose polymer. (b) Cellulose digestion sites by endoglucanase (EG) and cellobiohydrolase (CBH). Endoglucanase may digest glucose-glucose bonds along the entire polymer, except for the two outermost bonds at each end. Cellobiohydrolase only cuts off cellobiose from either polymer end. 
(c) Cellobiose digestion site by \textbeta -glucosidase (BGL). \textbeta -glucosidase exclusively digests cellobiose into two glucose molecules.
}
  \label{cellulase_action}     	
\end{figure*}

\noindent
 In this study, our main interest is in the production of glucose from a lignocellulosic substrate. Thus, the model includes the detail of the cellulases, which are responsible for the digestion of cellulose: endoglucanase (EG), cellobiohydrolase (CBH), and \textbeta -glucosidase (BGL). However, we implement the action of xylanases (XYL) in a coarse-grained fashion, i.e., a single enzyme that represents a cocktail of the xylanase sub-types. 
 In the model, EG, CBH, BGL and XYL are strictly distinct with regard to their mechanism, kinetics and abundance (see Figure \ref{cellulase_action}). While their specific mechanism is fixed in the model, their kinetics and abundance can be tuned. Also, we assume that EG mainly acts within the bulk of cellulose polymers, and therefore cannot digest the two outer most bonds on each end of the polymer. This implies that only CBH can release cellobiose. In addition, we omit the processive action of CBH, and instead implement the four enzymes as diffusive. Finally, we assume that the enzymes are homogeneously distributed within the medium over the course of the entire simulation.\\
 
 \noindent

\subsection{Substrate induced effects}\label{Saccha_challenges}

\noindent Many substrate induced effects are known to influence the efficiency of the saccharification process. The most important ones, which are also considered in our model, are briefly discussed below.\\

\noindent \textbf{Lignin adhesion.} Lignin acts as an adhesive towards both cellulases and xylanases, and is thought to negatively affect their action \cite{Ying2018, Weijde2016}. In the model, we represent the adhesive effect of lignin as an additional event that consists in binding a single randomly selected enzyme. The lignin surface available for binding is proportionally reduced such that 8 to 16 lignin monomers are now covered by the enzyme. Since the adhesive effect directly depends on the available lignin surface, lignin cannot bind an infinite number of enzymes. However, if the amount of lignin in the system is in large excess as compared to the enzymes, they can potentially all be bound, in which case the saccharification process is stopped. Also, the probability of an enzyme to bind to lignin depends on its relative concentration. This means that an enzyme whose concentration is lower than that of the others will be selected for adhesion to lignin less often, and \textit{vice versa}.\\

\noindent \textbf{Structural blocking.} Hemicellulose and lignin partly block the access of cellulases to the cellulose simply by their presence as a physical barrier around the microfibril \cite{Bura2009}. Therefore, the arrangement of hemicellulose and lignin has an impact on the saccharification by the cellulases. This is reinforced for lignin, since contrarily to hemicellulose it cannot be digested by the enzymes we consider. In the model, we assume that a cellulose bond covered by lignin cannot be digested until an adjacent bond has been digested, providing access to it for an enzyme. A bond covered by hemicellulose can be digested as soon as the hemicellulose is digested (see Figure \ref{accessibility_of_bonds}).\\

\noindent \textbf{Crystallinity.} It is well known that cellulose adopts a highly crystalline arrangement \cite{Park2010}, and that cellulose microfibrils contain crystalline and amorphous regions \cite{Harris2012}. While hemicellulose polymers are generally arranged in an amorphous manner \cite{Pauly2013}, to not restrict this aspect, we choose to consider both amorphous and crystalline regions for hemicellulose too. This is supported by the fact that we currently only consider xylan in our model, and xylan has been shown to partially bind to cellulose microfibrils, thereby adopting a semi-crystalline arrangement \cite{Simmons2016}. In the model, we therefore implement every cellulose and hemicellulose polymer as split into difficult to digest "crystalline" regions and more easily digestible "amorphous" regions. The size of the regions may be individually specified for cellulose and hemicellulose, as fractions of the overall cellulose and hemicellulose content. Additionally, $r_\textrm{c,a}$ denotes the ratio of the digestibilities for the crystalline regions ($d_\textrm{crystalline}$) \textit{versus} amorphous regions ($d_\textrm{amorphous}$):

\vspace{.5cm}

\begin{equation}
 r_\textrm{c,a} = \frac{d_\textrm{crystalline}}{d_\textrm{amorphous}}.
 \label{r_cry_amorph_equation}
\end{equation}

\vspace{.5cm}

\noindent It is a parameter that may also be tuned for cellulose and hemicellulose individually. It determines, how often a bond in a crystalline region is digested as compared to an amorphous region. Since crystalline regions are more difficult to digest than amorphous ones, $r_\textrm{c,a}$ lies between 0 (the crystalline region cannot be digested) and 1 (it is equally well digested as the amorphous region).

\subsection{A quick reminder of the Gillespie algorithm}

\noindent The saccharification of the substrate is simulated using a Gillespie algorithm. This is a typical method to implement stochastic simulations, and thereby mimic the dynamics of a system by assuming a sequence of randomized events \cite{Gillespie1976}. Here, these events correspond to reactions of enzymatic digestion. In the Gillespie algorithm we keep track of all and only doable reactions, by only accounting for the bonds accessible to the enzymes. At each step of the sequence of events, both the reaction to take place and its duration are randomly selected. Still, the algorithm ensures that the chance to select a reaction is on average equal to its likelihood to take place. The more reactions may take place, the more frequently an event happens. As an example, considering an available substrate in excess, more events take place per unit of time if the concentration of enzymes increases. The simulation lasts until the chosen amount of events is attained, or until all of the digestible substrate has been digested.\\

\noindent


\section{Results}\label{Model_analysis}

\noindent Our aim is to understand the effect of the substrate structure on the action of the enzymes as well as their interaction with non-digestible lignocellulose components. To do so,  we start from an investigation of the main features of the model. These are the synergism of the enzymes, the influence of lignin, and the impact of the crystallinity. Afterwards, we reproduce experimental saccharification time courses by Bura et al. \cite{Bura2009} for different pre-treatment conditions. We semi-quantitatively determine the impact of the substrate characteristics in shaping the saccharification process. Alongside this, we also interpret how pre-treatments affect the substrate structure. \\

\noindent \textbf{Default simulation conditions.} Unless otherwise specified, the simulations presented here were performed at the following default conditions: i) The microfibril length was set to 100 bonds, and the enzyme concentrations were each set to values of $[E] = $\,100 (arbitrary units). The enzyme concentration and the enzyme number are equal within the simulation, since we choose to set our volume as unity. This means that for a specified concentration $[E] = 100$ (arbitrary units), 100 enzymes are in the volume surrounding the microfibril. We ensure that such an enzyme number is reasonable by interpreting it for the experimental setup used by Bura et al. \cite{Bura2009} (see detailed calculation in Appendix \ref{Enzyme concentration}). We find that $ c_\textrm{enzyme} \approx 10\,\textrm{\textmu M},$

which lies within the range of some known \textit{in vivo} enzyme concentrations \cite{Albe1990}. ii) The corresponding rate of reaction of each enzyme is set to a default value of 1 (arbitrary units) if the enzyme is active, and 0 if it is not. iii) The default composition of the simulated microfibril is 50\,\% cellulose, 25\,\% hemicellulose, and 25\,\% lignin. We choose these as convenient values which fall well into the ranges stated by Zoghlami and Pa\"{e}s \cite{Zoghlami2019} (biomass weight is 40-60\,\% cellulose, 20-35\,\% hemicellulose, and 15-40\,\% lignin). iv) Either 100 or 1\,000 simulation runs are performed to study the average behavior of the system, which is by definition stochastic.\\

\subsection{Enzymatic synergism}

\begin{figure*}
  \centering
  \subfigure[Only EG present]{
    \includegraphics[width=0.3\textwidth]{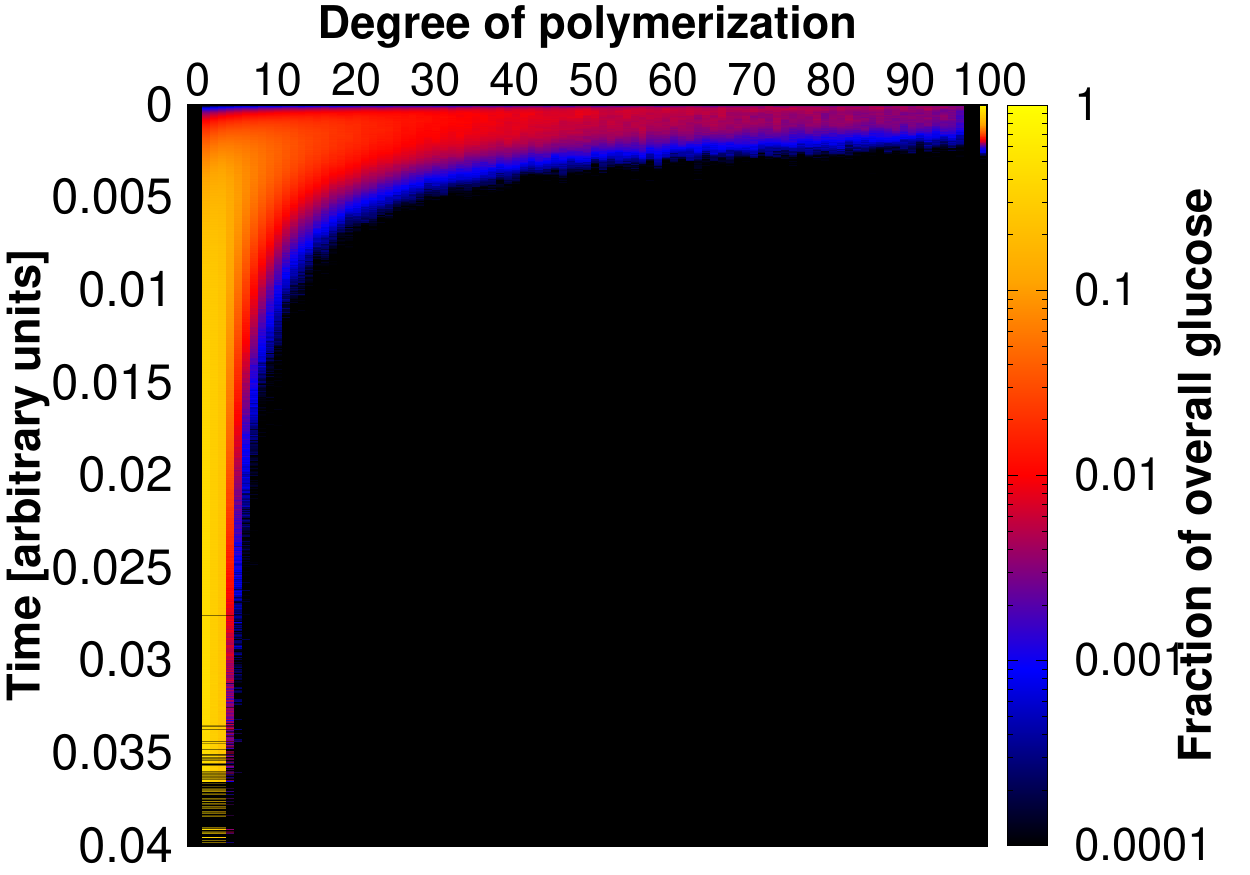}
  }
  \subfigure[Only CBH present]{
    \includegraphics[width=0.3\textwidth]{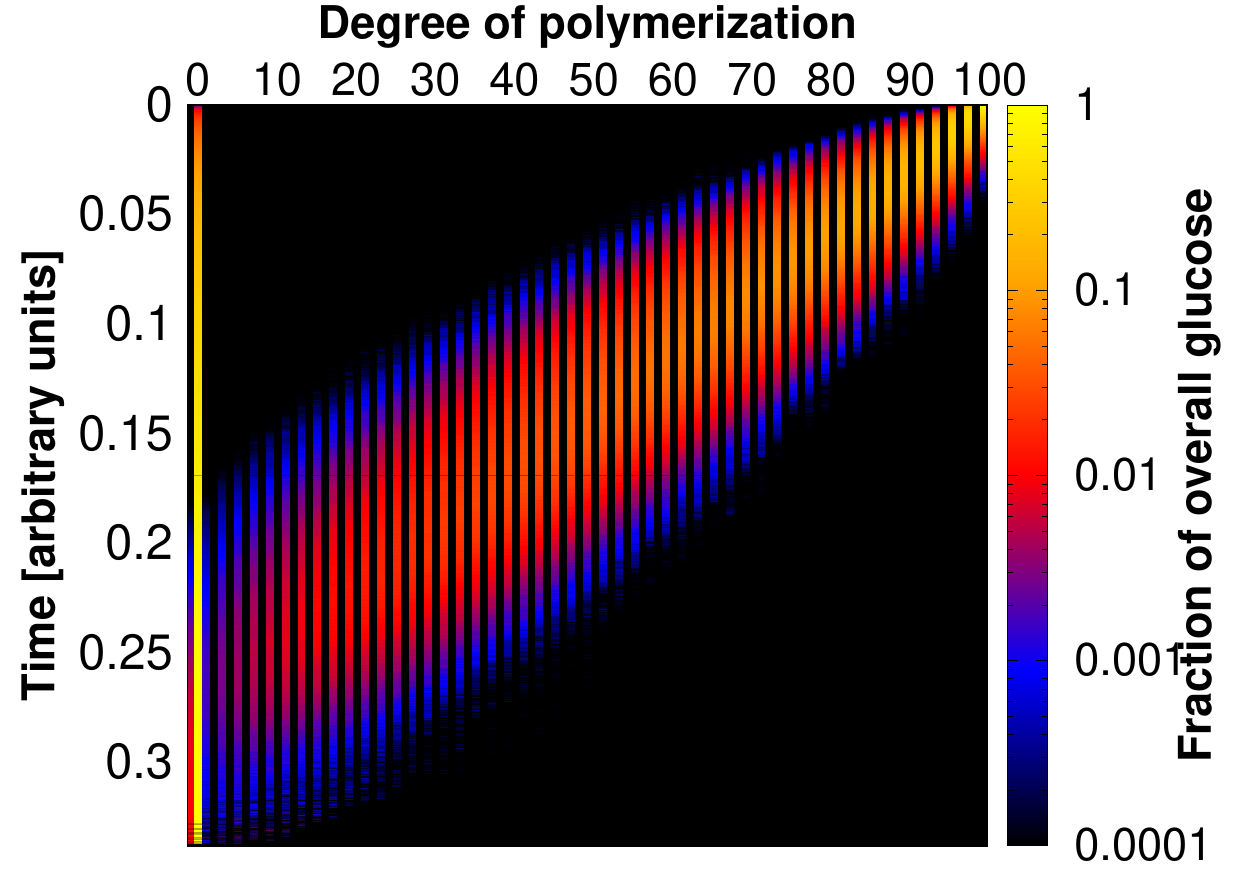}
  }
  \subfigure[EG, CBH and BGL present]{
    \includegraphics[width=0.3\textwidth]{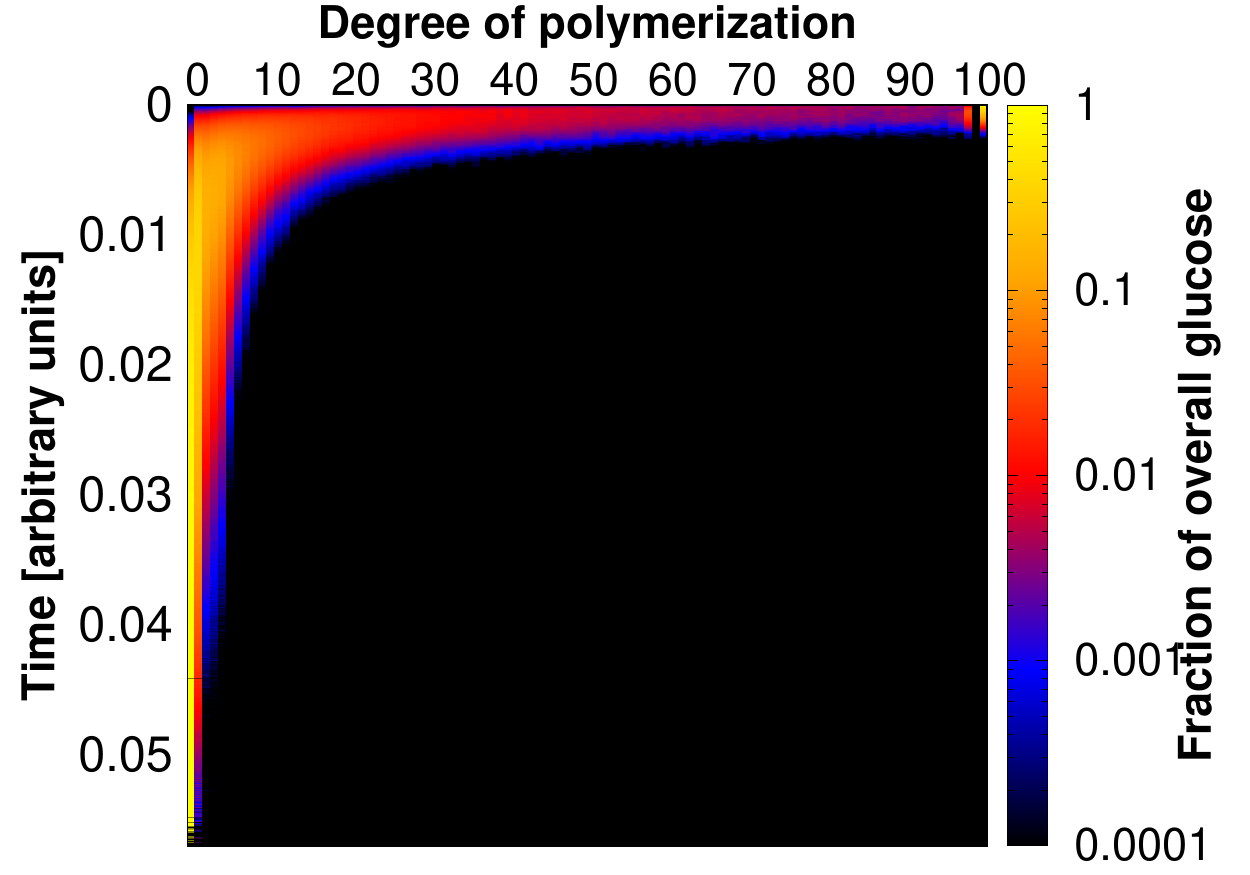}
  }
  \captionof{figure}{
 Heatmaps depicting the dynamics over time of the distribution of simulated cellulose polymer degrees of polymerization (DP). Time is shown on the ordinate, DP is shown on the abscissa, and color indicates the total amount of glucose that makes up polymers of length DP divided by the total amount of glucose in the system. Only present enzymes are: EG in (a), CBH in (b), and EG, CBH and BGL all together in (c). In (a), cellotriose accumulates, since in the model EG alone cannot release cellobiose or glucose. In (b) cellobiose progressively accumulates, while a small amount of glucose is produced. Finally, in (c) cellobiose can be digested by BGL, and in turn glucose accumulates. Each heatmap represents an average over 1\,000 simulation runs.
  } 
  \label{Heatmaps}     	
\end{figure*}

\noindent The enzyme cocktails used by plant-digesting fungi are effective only due to the combined action of the individual enzymes \cite{Joergensen2007}. Here we utilize the model to decipher the action of these enzymes individually, and then collectively, in order to explain and quantify their synergism. The heatmaps shown in Figure \ref{Heatmaps} depict the distribution of simulated cellulose polymer degrees of polymerization over time, for different enzyme sets. Since we only investigate in detail the synergistic action of the cellulases in that figure, the simulated microfibril is not surrounded by an outer shell of hemicellulose and lignin.\\

\noindent When only endoglucanase (EG) is present (subfigure (a)), the distribution of cellulose quickly changes, from initially being made of the longest polymers, to being split over the shorter ones. This can be explained by the fact that endoglucanase can randomly cut glycosidic bonds along cellulose filaments, which results in polymers of any length. However, since in the model EG does not cut off cellobiose or glucose from the polymers, none of these two are released, and the two leftmost columns of the heatmap remain black throughout the simulation. Instead, cellotriose accumulates over time (third column from left). Symmetrically, no polymers of length 99 or 98 appear (second and third column from the right).\\

\noindent When only cellobiohydrolase (CBH) is present (subfigure (b)), the distribution moves towards shorter polymers linearly in time, while cellobiose (second column from left) steadily increases. Starting from a monodisperse pool of chain length 100, the digestion by CBH, which can only release cellobiose, leads to only even chain lengths. This appears as vertical stripes in the distribution. The linearity of the profile emerges as CBH cuts off cellobiose from any of the two ends of the cellulose polymers, meaning that the number of available attack points for the enzyme remains equal over time. This is because no new polymer is created by the action of CBH, and each polymer systematically presents two digestible ends for CBH. The small amount of glucose (leftmost column), released towards the end of the simulation, results from the digestion of cellotriose (third column from left).\\

\noindent Importantly, the duration of the saccharification can be compared between subfigures, since the (arbitrary) time units are similar. It is apparent that the complete digestion by CBH (subfigure (b)) is slower than that by EG (subfigure (a)). Because here EG and CBH have the same concentration and kinetics, we can explain this by the number of attack points along the polymers for each of them, which directly depends on their mechanism. For instance, a fully accessible cellulose polymer of a length of 100 bonds represents 96 attack points for EG, while CBH can only attack each end.\\

\noindent Finally, in subfigure (c) we simulate the action of a set of cellulases that contain EG, CBH and BGL together. The distribution moves towards the shorter polymers rather similarly to the case of only EG being present (subfigure (a)). This indicates that at equal concentrations, the dynamics of the saccharification process is dominated by the enzyme that has the most numerous accessible attack points on the substrate, on the condition that these do not require the action of another enzyme to be made available. Additional features characterize the synergistic action of the distinct enzymes. Firstly, although the duration of the complete saccharification is approximately similar as in subfigure (a), the distribution shape moves towards short polymers in a steeper fashion. This can be explained by the fact that the action of EG creates new polymers, which each constitute two attack points for CBH. Secondly, the cellotriose, which accumulates via the action of EG only, is now digested by CBH into cellobiose and glucose. Thirdly, the cellobiose, which accumulates under the action of CBH only, is now digested into two glucose molecules due to BGL.\\

\subsection{Lignin influence}

\noindent The lignin adhesion properties have already been computationally investigated by Vermaas et al. \cite{Vermaas2015}. Using molecular dynamics simulations, they mimicked an atomic-detail cellulose microfibril surrounded by freely floating cellulases and lignin oligomers. They showed that the lignin oligomers in their system have a high affinity for binding both to the cellulose-binding domain of the investigated cellulase, and to those positions on the cellulose microfibril which are also preferred locations of action by the enzymes. The two effects exhibited by lignin within our model are in line with their conclusions. We consider the adhesion effect towards enzymes, and we additionally include the structural blocking effect resulting from the arrangement of the lignin around the microfibril. Furthermore, our model allows to investigate how these effects impact the saccharification curves arising from different lignin contents.\\ 

\begin{figure*}
  \centering
  \subfigure[]{
    \includegraphics[width=0.45\textwidth]{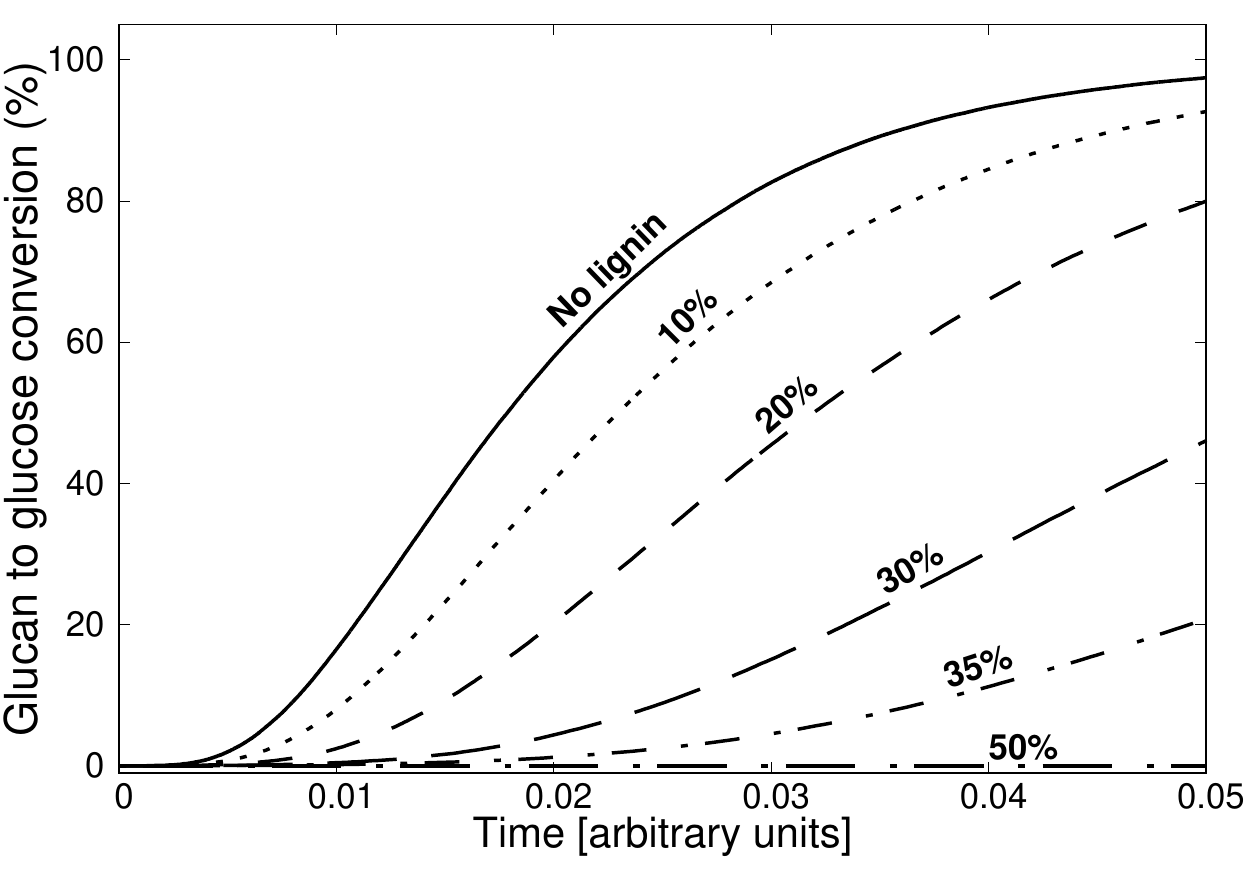}
  }
  \subfigure[]{
    \includegraphics[width=0.45\textwidth]{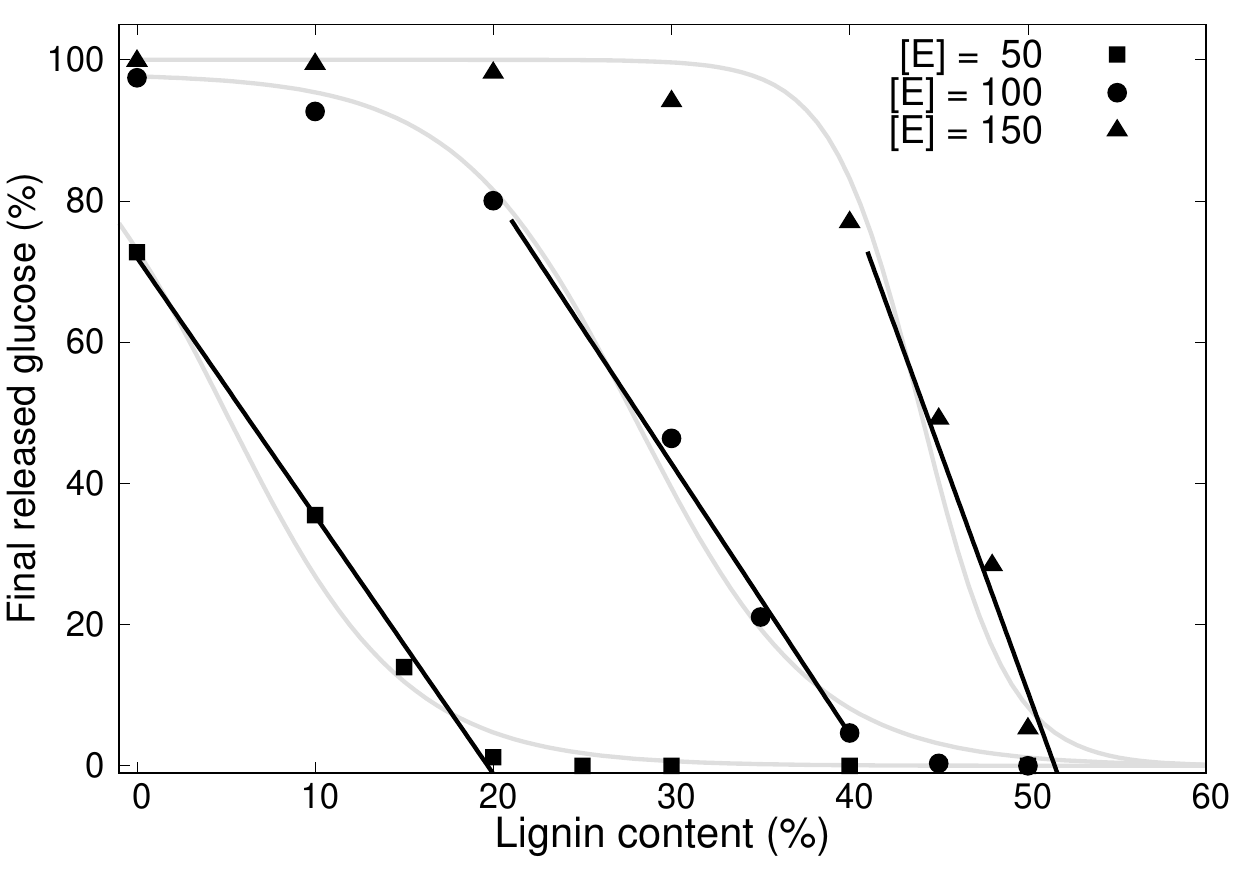}
  }
  \subfigure[]{
    \includegraphics[width=0.45\textwidth]{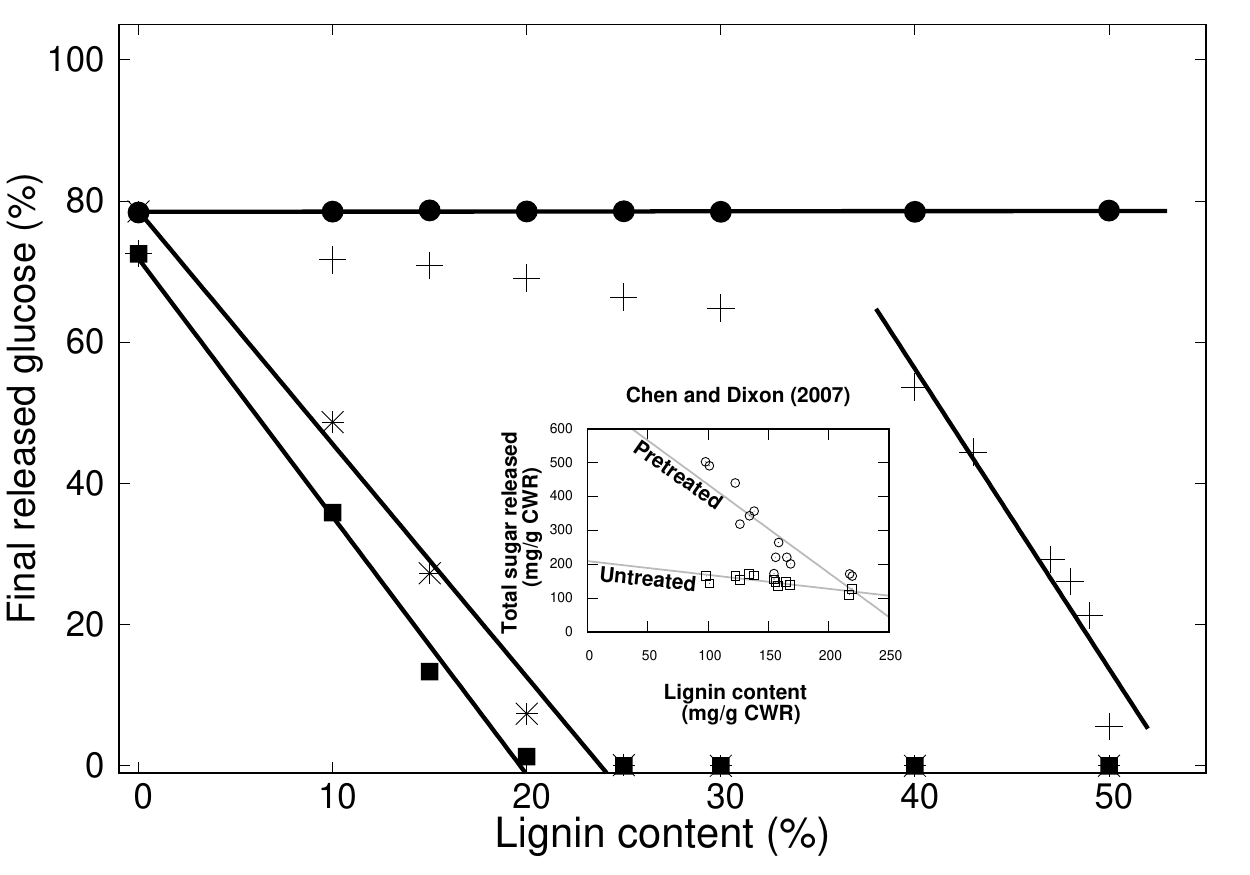}
  }

  \captionof{figure}{
  (a) Simulated cellulose to glucose conversion dynamics curves for increasing lignin percentage up to a time $t_\textrm{end} = 0.05$ (arbitrary units). Each curve represents an average over 100 simulation runs. (b) Simulated final cellulose to glucose conversion yield at time $t_\textrm{end}$ \textit{versus} lignin percentage for three different values of enzyme concentrations $[E]$. Also shown are trend lines for the approximately linear intermediate regimes. (c) Simulated final cellulose to glucose conversion yield at time $t_\textrm{end}$ \textit{versus} lignin percentage for different combinations of lignin effects. Four situations are shown: deactivated lignin adhesion and deactivated structural blocking (circles), deactivated lignin adhesion and activated structural blocking (crosses), activated lignin adhesion and deactivated structural blocking (stars), and finally both active (squares). Other simulation parameters are the same as for the case $[E] = 50$\,(arbitrary units) in subfigure (b). The situation in which both effects are active therefore corresponds to that in subfigure (b) (squares). Also shown are experimental data and linear regression curves by Chen and Dixon \cite{Chen2007}, who analyzed the saccharification yield of lignocellulose from alfalfa mutants, containing different amounts of lignin, both for acid-pretreated samples and untreated samples. (a), (b) and (c) The microfibril simulated is of length 200 bonds. (b) and (c) Each point represents an average over 100 simulation runs.}
  \label{lignin_percentage_vs_saccharification}     	
\end{figure*}

\noindent In Figure \ref{lignin_percentage_vs_saccharification} (a), we show the saccharification dynamics up to a time $t_\textrm{end} = 0.05$\,(arbitrary units), for a substrate of length 200 bonds. This is composed of 50\,\% cellulose, with a varying outer hemicellulose and lignin shell, such that the overall lignin percentage ranges from 0 to 50\,\%. At $t = t_\textrm{end}$, the cellulose within the substrate containing no lignin is almost completely digested. At increasing lignin percentage, the digestion of the substrate takes longer. We observe a steeper decrease in saccharification yield at $t = t_\textrm{end}$ for the same increase in lignin percentage at higher overall lignin content. This indicates, that within our model the dependence of the saccharification yield on the lignin content is non-linear. Upon investigating only the yields at $t = t_\textrm{end}$, for three different enzyme concentration values (subfigure (b)), we see that overall the change in yield behaves similarly to a logistic decay (gray lines). This is characterized by an initially small slope, followed by an intermediate strong decay towards a yield of 0.  The behavior within the domain of strong decay can also be approximated as linear, indicated by the trend lines.\\

\noindent By comparing the three curves in subfigure (b), we see that the overall behavior of the yields at time $t_\textrm{end} = 0.05$\,(arbitrary units) depends strongly on the ratio ($r_\textrm{E,L}$) between the enzyme concentration and the lignin content. This is due to the fact that only a finite number of enzymes can bind to a given amount of lignin. Considering for instance the rightmost curve (triangles, [E] = 150 (arbitrary units)), at high $r_\textrm{E,L}$ (leftmost points of the curve) the lignin negligibly influences the action of the enzymes and we observe a plateau. At low $r_\textrm{E,L}$ (rightmost points of the curve), the lignin strongly disrupts the saccharification yields, and we observe a sharp drop of the curve. By reducing the enzyme concentration (other two curves), this typical profile is shifted to the left, such that the strong disruption regime appears at lower lignin content. The trade-off of enzymes that either perform saccharification, or get inactivated by binding to lignin, underpins the logistic decay behavior.\\

\noindent To further investigate the action of lignin, we separately examine the adhesion effect and the structural blocking effect. Within subfigure (c) we show four different simulation results, in which each effect is either active or inactive. Other simulation parameters are the same as for the case $[E] = 50$\,(arbitrary units) in subfigure (b). The situation in which both effects are active therefore corresponds to that in subfigure (b) (squares). Also shown are experimental data by Chen and Dixon \cite{Chen2007}. They analyzed the saccharification yield of lignocellulose from alfalfa mutants containing different amounts of lignin, both for acid-pretreated samples and untreated samples. We show the linear regression curves they provide together with their data.\\

\noindent In the control situation (circles), when both effects are inactive, lignin, as we expect, has no impact on the saccharification yield, which therefore remains constant. When only the structural blocking effect is active (crosses), we see a slow decrease in final saccharification yield, which starts to decay strongly at high lignin percentage (between 40\,\% and 50\,\%), when the outer shell is almost completely made of lignin. As the microfibril is increasingly surrounded by indigestible lignin, there are less bonds which can be accessed by enzymes, and the relative decrease in accessible bonds is stronger at high lignin percentage than at low lignin percentage. At high lignin percentage, only CBH and BGL can access digestible bonds, while at low lignin percentage the bulk of the cellulose polymers can be uncovered by lignin, and thus EG also can access digestible bonds. On the other hand, when only the adhesion effect is active (stars), we observe an immediate drop in final saccharification yield for increasing lignin percentage, similar to that observed in subfigure (b). Since we use low enzyme concentration ($[E] = 50$\,(arbitrary units)), the amount of lignin in the system is quickly sufficient to significantly deplete enzymes and to make the trade-off between adhesion and saccharification visible. Finally, the situation in which both effects are active (squares) displays their cumulative behavior.\\

\noindent In subfigures (b) and (c) we include trend lines for comparison with experimental measurements. In the inlay of subfigure (c), we show Chen and Dixon's data for untreated and acid pre-treated samples, as well as the linear regression between lignin content and saccharification yield they obtained for both situations \cite{Chen2007}. A similar linear decrease tendency has been observed by Studer et al. \cite{Studer2011} for distinct plant materials. They selected different phenotypes of poplar trees based on their lignin content, and also studied their saccharification dynamics. With our model, we are able to show a comparable linear decrease, which supports that the mechanisms we consider for lignin, in terms of its interactions with enzymes and the blocking of the structure, fairly represent it. Importantly, our model offers to investigate the dynamics of these effects throughout the saccharification process and, being a flexible theoretical tool, it allows us to study ranges of lignin content beyond the typical experimental results found in the literature. Screening a larger range of lignin percentages reveals that the experimentally observed linear decrease can possibly be interpreted as the intermediate regime of a logistic decay, which also shows a plateau-like profile at low lignin content.

\subsection{Substrate crystallinity}\label{crystallinity_section}

The impact of cellulose crystallinity on the saccharification yield is a matter of recent focus \cite{Pena2019,Cui2014}. To reflect it in the model, we split the cellulose and hemicellulose into difficult to digest "crystalline" regions and easy to digest "amorphous" regions. The likelihood for digestion of polymer bonds within amorphous regions is higher than that of polymer bonds in crystalline regions. To investigate the influence of the crystallinity, we consider three sets of simulations. The difference between them lies in the ratio ($r_\textrm{c,a}$) between the digestibilities of the amorphous and crystalline regions. In each set, the fraction of crystalline regions varies between 0 and 90\,\% of the respective substrate, and the resulting final saccharification yields are compared. The latter are defined as the yields obtained at a specific time $t_\textrm{end}$, that we choose as $t_\textrm{end} = 72$ (arbitrary units) for later comparison with experimental data (see section \ref{Model_comparison}). In each respective simulation the cellulose and hemicellulose crystallinity parameters were set equal to each other. In Figure \ref{Various_crystallinities_figure}, we observe a linear decrease of the final saccharification yields with the increase in crystallinity fractions. The corresponding slopes become steeper at lower $r_\textrm{c,a}$. As expected, we observe inverse proportionality in addition to linearity when $r_\textrm{c,a} = 0$, i.e., when the crystalline regions are completely indigestible.\\  

\noindent Also shown in Figure \ref{Various_crystallinities_figure} are experimental data by Cui et al. \cite{Cui2014} and Pena et al. \cite{Pena2019}. Cui et al. investigated the influence of four different pre-treatment methods (ionic liquid, ethylenediamine, glycerol, and sodium hydroxide) on the crystallinity of \textalpha -cellulose samples, and further investigated their saccharification yield depending on the crystallinity. They observed an inverse linear relationship between the measured crystallinity index and the saccharification yield. Pena et al. on the other hand focused exclusively on the ionic liquid pre-treatment method, and investigated the cellulose crystallinity after different pre-treatment durations. Their measurement of the saccharification yield also shows an inverse linear relationship. Importantly, both sets of data are reasonably well matched by the model, even though the experimental setup varies considerably between them. We remark that the curve by Cui et al. is shifted to the right, with complete glucose conversion achievable despite about $22\%$ cellulose crystallinity. This might suggest a slightly different definition of the crystallinity index.

\begin{Figure}
  \centering

    \includegraphics[width=\textwidth]{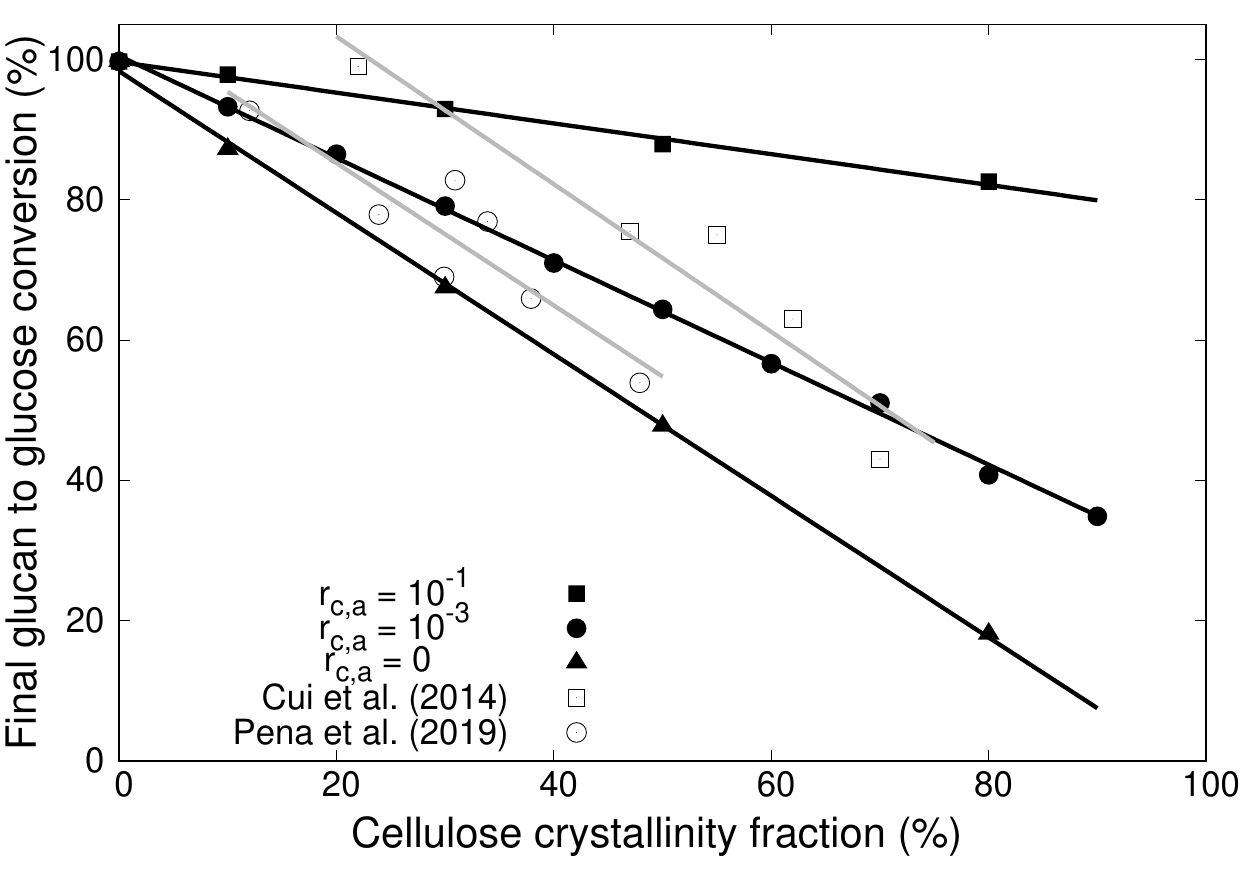}

  \captionof{figure}{
 Simulated final saccharification yield \textit{versus} crystallinity fraction for different ratios ($r_\textrm{c,a}$) between the digestibilities of the crystalline and the amorphous regions. For each value of $r_\textrm{c,a}$ we observe a linear decrease in saccharification yield for increasing crystallinity fraction, whose slope becomes steeper as $r_\textrm{c,a}$ decreases. For $r_\textrm{c,a} = 0$ we observe inverse proportionality in addition to linearity. Each simulated point represents an average over 100 simulations. Also shown are experimental data by Cui et al. \cite{Cui2014} (empty squares) and Pena et al. \cite{Pena2019} (empty circles).
} 
  \label{Various_crystallinities_figure}
\end{Figure}

\subsection{Comparison of the model to experimental data for pre-treated plant cell wall material}\label{Model_comparison}

\noindent \textbf{Fitting procedure.} In order to investigate the capability of the model to precisely reproduce experimental saccharification time courses, we devised a parameter fitting algorithm. It can be applied to any parameter of the model to optimize it. Among the set of all the parameters which characterize the model, we choose to optimize only a subset of parameters $p_\textrm{i}$. The algorithm works in recursive generations, that each build on the preceding one. Each generation contains a number of distinct subsets, whose parameters vary within a percentage $\Delta$ from the preceding generation of parameters ($p_\textrm{i, old}$). More precisely, each subset's parameters $p_\textrm{i}$ are randomized, but lie within the respective interval $[p_\textrm{i, old} - \Delta \cdot p_\textrm{i, old}; p_\textrm{i, old} + \Delta \cdot p_\textrm{i, old}]$. The total number of subsets within a single generation can be freely specified at the start of the algorithm.\\

\noindent At each generation, the saccharification process is simulated for all the subsets of parameters, and the glucose and the xylose yields are recorded. Then, the difference between the simulated and the experimental results is measured for successive time points along the saccharification curves. This is done for glucose and xylose curves separately, and can be applied to multiple experimental datasets simultaneously, for instance different pre-treatments. The resulting errors are then combined within an average error between theory and experiments, which depends on all different glucose and xylose curves.\\ 

\noindent After the error between simulated and experimental curves has been measured for each subset of parameters, the best fitting candidate, with the lowest error, is found. If this error is higher than that of the subset of parameters $p_\textrm{i, old}$, the old parameters are kept and used again as a starting point for the subsequent generation. Otherwise, the parameters of the new subset are retained. Additionally, the direction within the parameter space between the old and the new parameters is calculated. This direction is followed for a number of subsequent generations. If the error has not been reduced after that number of generations, the algorithm reverts to randomly assigning parameter values depending on $\Delta$. After multiple generations, this hybrid procedure that mixes random and directed search leads to a local or global optimum in precision.\\

\begin{figure*}
  \centering
  \subfigure[]{
    \includegraphics[width=0.3\textwidth]{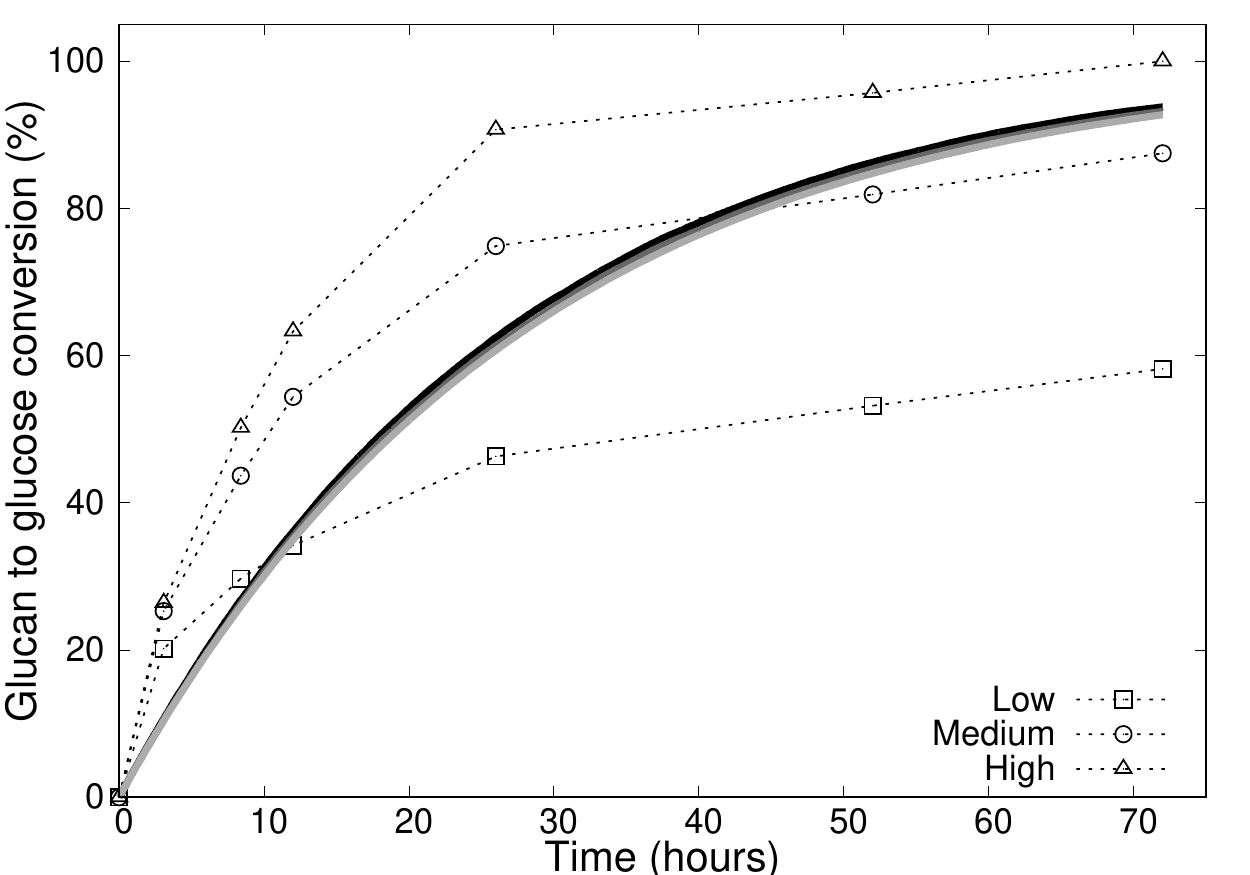}
  }
  \subfigure[]{
    \includegraphics[width=0.3\textwidth]{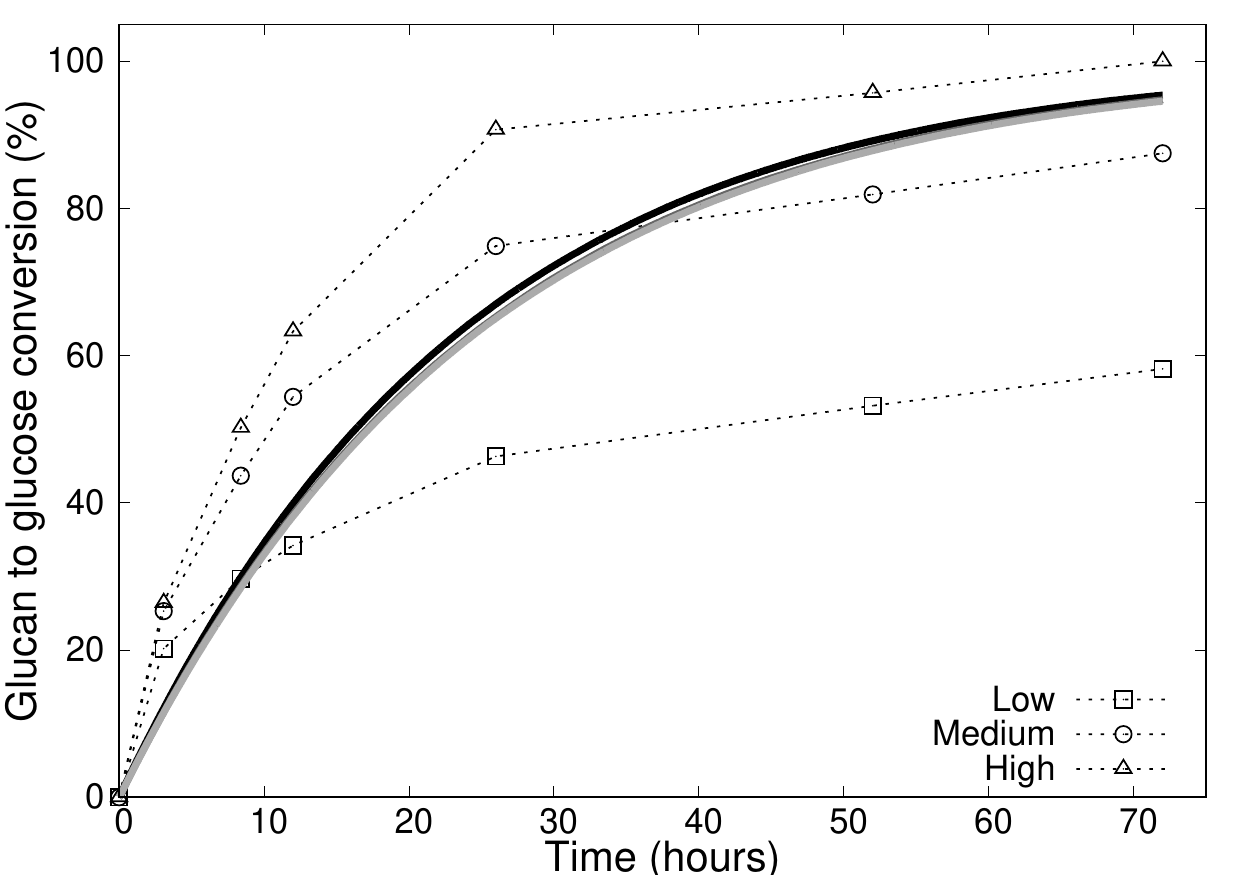}
  }
  \subfigure[]{
    \includegraphics[width=0.3\textwidth]{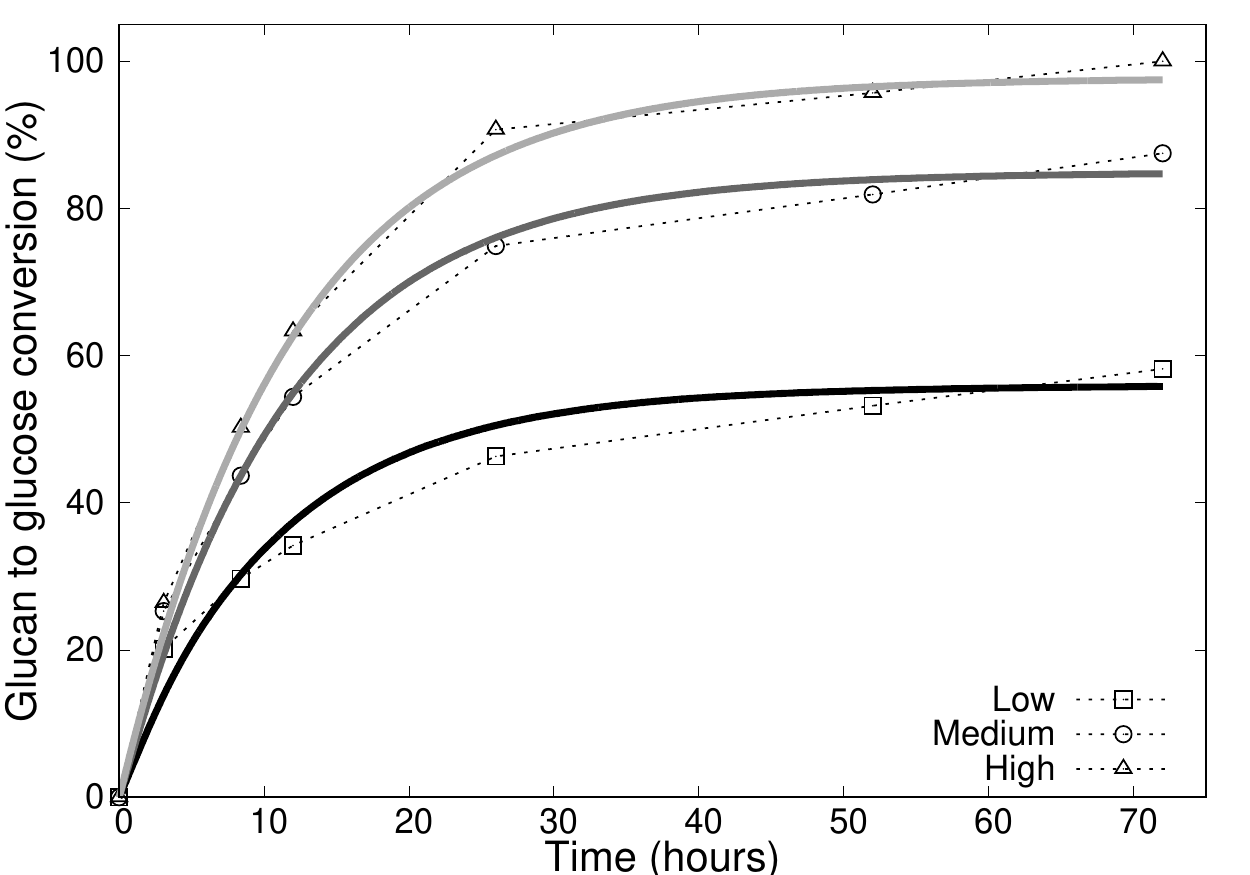}
  }  
  \subfigure[]{
    \includegraphics[width=0.3\textwidth]{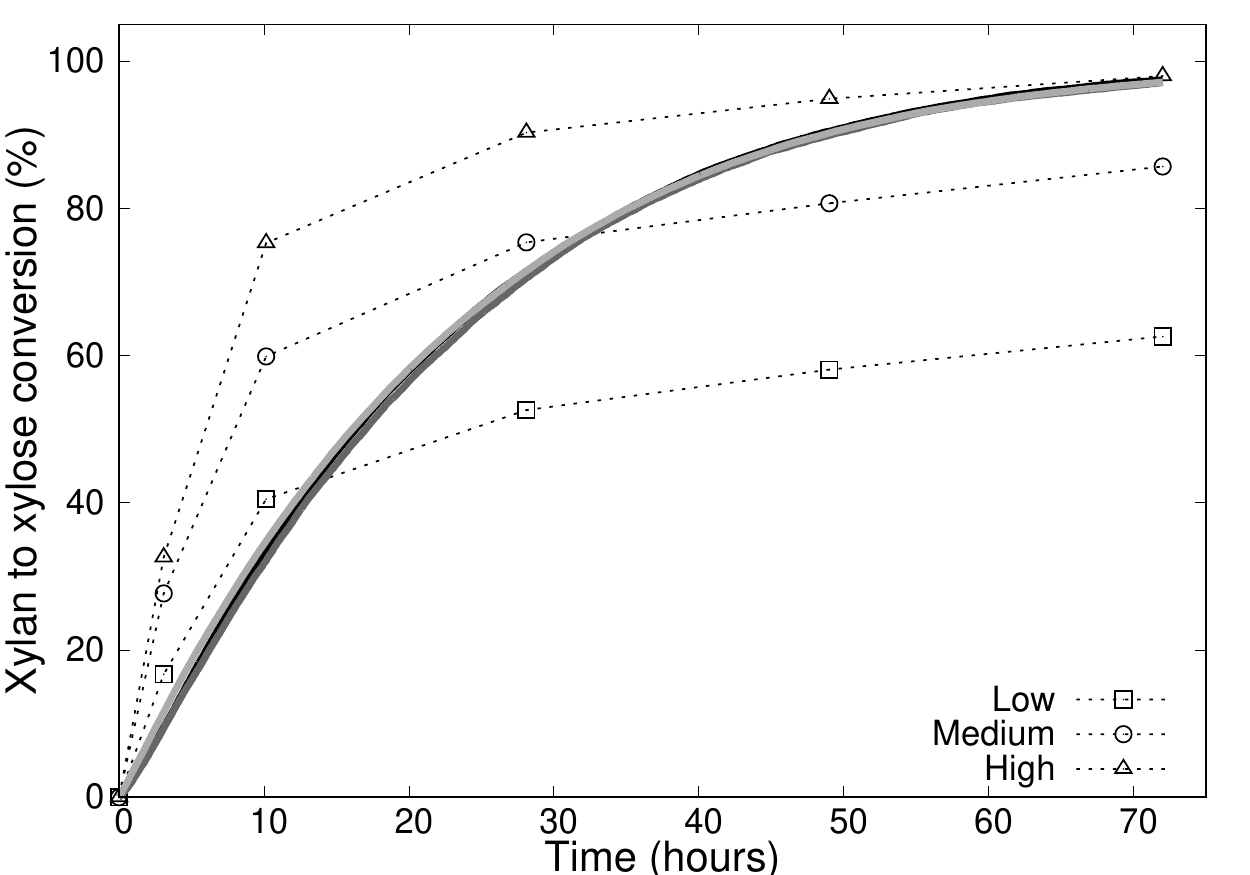}
  }  
  \subfigure[]{
    \includegraphics[width=0.3\textwidth]{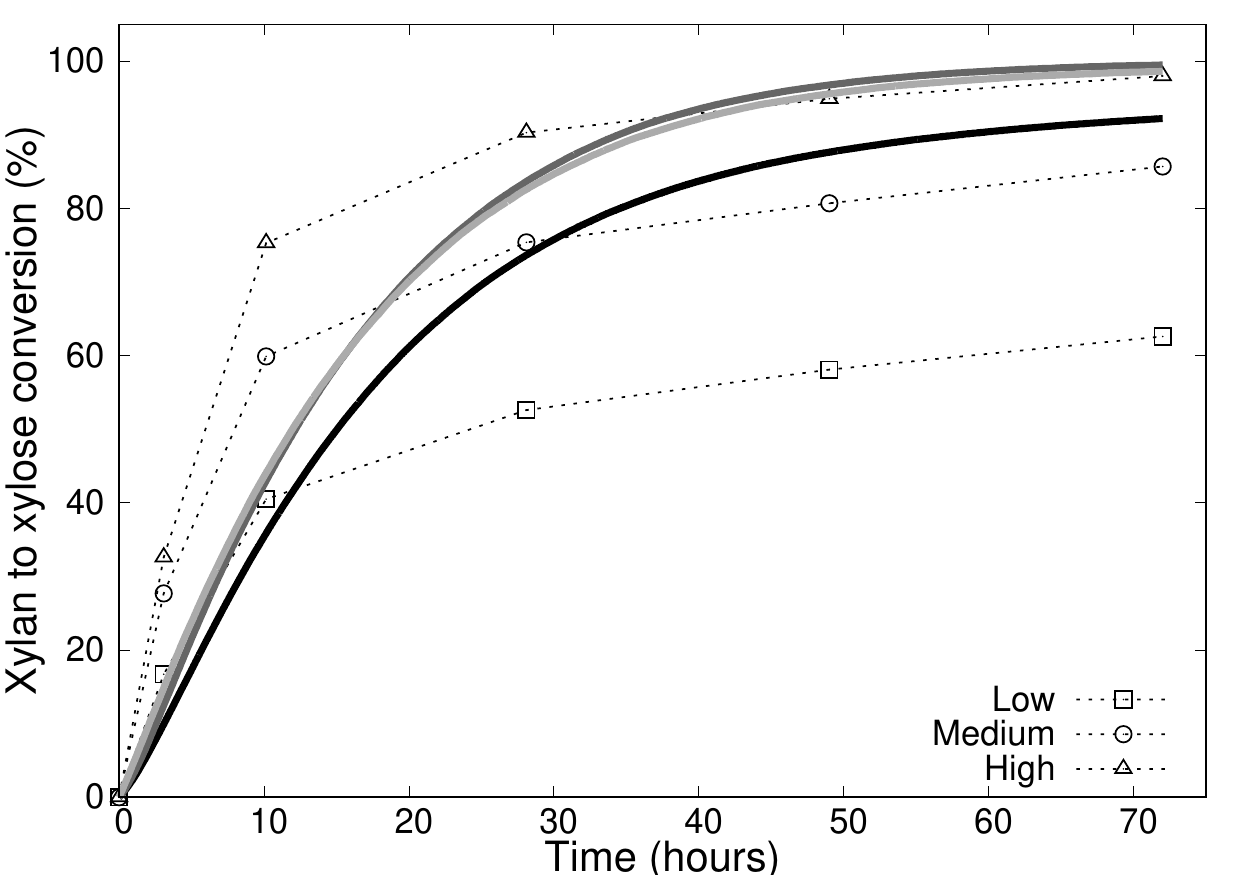}
  }
  \subfigure[]{
    \includegraphics[width=0.3\textwidth]{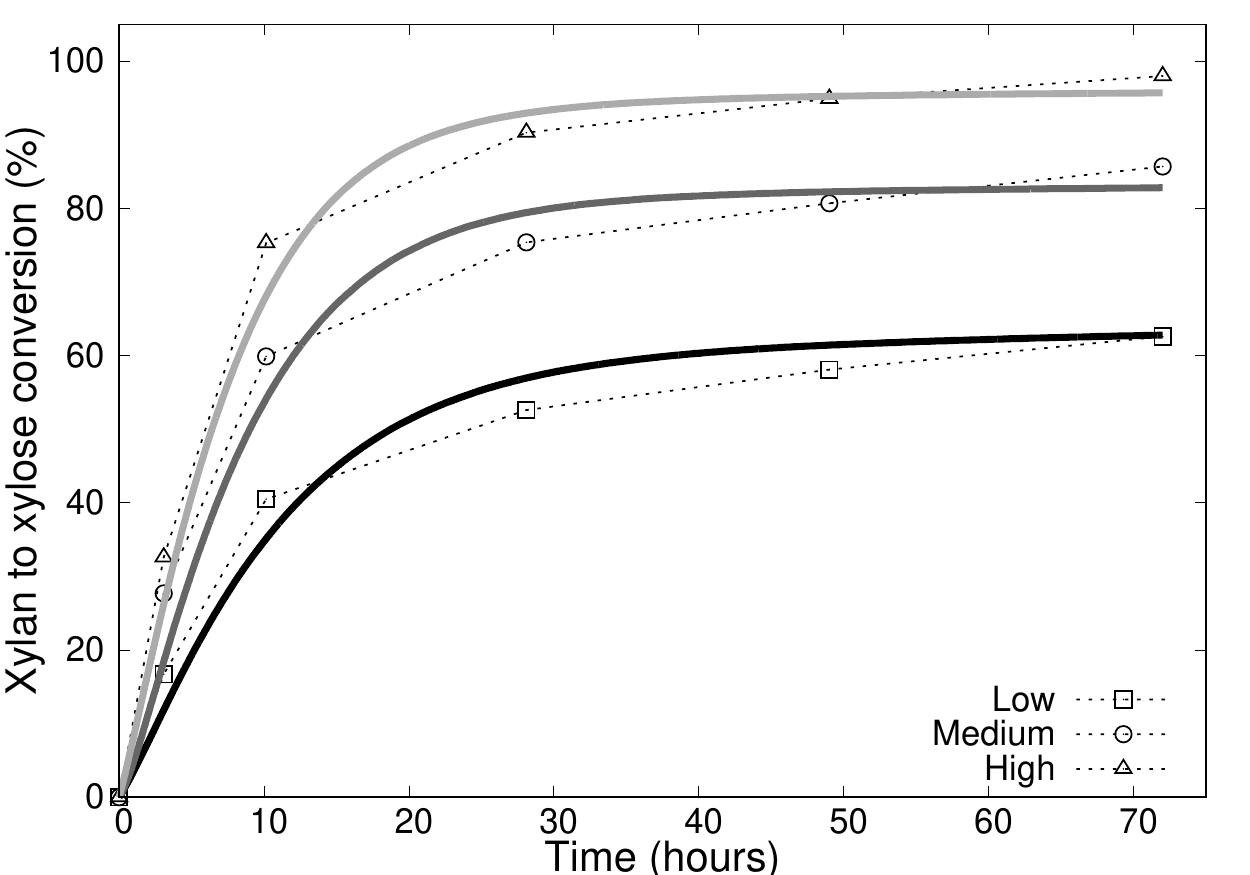}
  }
  \captionof{figure}{
Experimental data (dotted lines) and best simulation fits of the saccharification time courses for three different pre-treatment severities (low: black lines, medium: dark grey lines, high: light grey lines). (a) and (d) The substrate has no structure and all polymers are freely floating within the medium. (b) and (e) The cellulose polymers form a microfibril, which is surrounded by hemicellulose and lignin. However, the crystallinity of the substrate is discarded. (c) and (f) The substrate crystallinity is additionaly included, which substantially improves the fits. Each curve represents an average over 1\,000 simulation runs. The experimental data shown here are from Bura et al. \cite{Bura2009}. 
} 

  \label{Bura_comparison}
\end{figure*}

\noindent \textbf{Experimental approach.} To demonstrate the ability of our model to reproduce experimental saccharification time courses, we investigate results by Bura et al. \cite{Bura2009}. They studied the influence of a steam pre-treatment process on the saccharification yield of plant material from corn stover. For this, they first subjected their samples to three different pre-treatment severities, denoted "low", "medium" and "high". The severities were characterized both by the temperature during the pre-treatment and the overall duration of the pre-treatment. Following the pre-treatment, the samples were examined with respect to their composition of cellulose, hemicellulose and lignin. Table \ref{Param_table} outlines the substrate compositions resulting from the three pre-treatment severities. Then, a cocktail of cellulases and xylanases was added to the samples, and the glucan (cellulose) to glucose and xylan (hemicellulose) to xylose conversion percentages were periodically measured over 72 hours (see dotted lines in Figure \ref{Bura_comparison}). Higher saccharification yields were observed for higher pre-treatment severities, and it was concluded that the concomitant reduction of the xylan content was the main influence in determining the saccharification yield.\\

\noindent \textbf{Simulation results.} The findings of Bura et al. suggest that the substrate composition in terms of cellulose, hemicellulose and lignin determines the saccharification yield. We investigate this within our model by using the composition data provided by them and attempting to optimize the parameters of the model to reproduce the experimental saccharification curves. Our goal is to find a set of parameters with which all six experimental curves (glucose conversion and xylose conversion for each of the three pre-treatment severities) can be reproduced by only adapting the substrate composition to the respective pre-treatment conditions (see Table \ref{Param_table}). Starting from the hypothesis that the composition of the substrate solely influences the yield, we progressively additionally consider the effect of its structural properties. Thus, in the following we consider three situations, and we use the fitting procedure to optimize the enzyme rates of reaction and the rate of adhesion to lignin in each case. In the third case, we additionally optimize the crystallinity fractions and the ratio between the digestibilities of crystalline and amorphous regions. The simulated saccharification time courses for the three situations investigated are shown with the experimental data in Figure \ref{Bura_comparison}. \\ 

\noindent In subfigures (a) and (d), we simulate a substrate in which all polymers are assumed to freely float in the medium instead of being arranged within a microfibril. Therefore, every digestible bond (both cellulose and hemicellulose) is accessible from the beginning of the simulation. As can be seen, it is not possible for us to find a set of parameters capable of accurately reproducing the data for this situation. In subfigures (b) and (e), we consider a spatially structured microfibril made of 24 cellulose chains and surrounded by hemicellulose and lignin like in Figure \ref{Microfibril_structure}. However, crystallinity properties are so far discarded. Such hypotheses do not yield satisfying results either. The best fitting glucose conversion curves (solid lines in subfigure (b)) are almost identical for all three pre-treatment severities, and are closest to the data for medium pre-treatment. For the xylose conversion curves (solid lines in subfigure (e)), the low pre-treatment severity can be distinguished from the other two, and is the lowest. However, both medium and high pre-treatment severities cannot be clearly distinguished, and none of these three lies close to the respective experimental data. Therefore, the substrate composition alone does not enable reproducing and explaining the experimental data, and neither does considering the substrate structure without distinguishing between crystalline and amorphous regions.\\

\noindent To elucidate this problem, we incorporate more advanced structural properties of the substrate, and so the crystallinity is included in the third case. The crystallinity fractions and the ratio between the digestibilities of crystalline and amorphous regions ($r_\textrm{c,a}$) are optimized, and can be found in Table \ref{Param_table}. As can be seen in subfigures (c) and (f), the agreement between simulations and experiments indeed improves substantially. The simulated glucose conversion curves (solid lines in subfigure (c)) fit the experimental data well, as do those of xylose (solid lines in subfigure (f)). The crystallinity fractions found decrease drastically with increasing pre-treatment severities. This makes sense if one considers the different temperatures used within the three pre-treatment severities. The order of crystalline structures is generally reduced for increasing temperature, and therefore a higher severity leading to a reduction in crystallinity fraction is plausible. In addition, as discussed in section \ref{crystallinity_section}, it has already been demonstrated experimentally that the cellulose crystallinity influences the saccharification yield \cite{Pena2019,Cui2014}. Importantly, the results also support the hypothesis by Simmons et al. \cite{Simmons2016} that xylan adopts a semi-crystalline shape around cellulose microfibrils. In fact, the inclusion of both cellulose and hemicellulose crystallinity is necessary for the fits of the experimental data to be optimal.\\

\noindent Within the simulation scheme, the results imply a significant influence of the crystallinity on the saccharification dynamics and final yield. Furthermore, the crystallinity allows to both reproduce accurately the experimental data and to explain the differences between the yields for the different pre-treatment severities.\\

\begin{table*}
\centering
\begin{tabular}{|c|l|l|l||l|l|l|l|}
  \hline
  \multirow{2}{*}{\textbf{Pre-treatment severity}} 
      & \multicolumn{3}{c||}{\textbf{Composition (\%)}} 
          & \multicolumn{2}{c|}{\textbf{Crystallinity fractions (\%)}}
       & \multicolumn{2}{c|}{\textbf{$r_\textrm{c,a}$ (\%)}} \\ \cline{2-8}
  & Cellu & Hemi & Lignin & Cellu & Hemi & Cellu & Hemi\\  \hline
  Low & 57.0 & 18.8 & 24.2 & 53 & 47 & 0.16 & 0.73\\      \hline
  Medium & 61.9 & 9.5 & 28.6 & 21 & 26 & 0.16 & 0.73\\      \hline
  High & 62.7 & 5.3 & 31.1 & 2.6 & 6.2 & 0.16 & 0.73\\      \hline  
\end{tabular}
\captionof{table}{
Parameters for the substrate composition, crystallinity fractions, and digestibilities ($r_\textrm{c,a}$) at three pre-treatment severities (low, medium and high). The composition is closely derived from experimental data: the percentages of hemicellulose and lignin are from Bura et al. \cite{Bura2009}, while the glucose percentages are adapted such that the composition percentages sum up to 1. The crystallinity fractions and digestibility ratios $r_\textrm{c,a}$ (see equation (\ref{r_cry_amorph_equation})) are determined by fitting the experimental saccharification time courses by Bura et al. shown in Figure \ref{Bura_comparison}.
}
\label{Param_table}
\end{table*}

\section{Discussion and conclusions}\label{discussion_conclusion}

\noindent The enzymatic digestion and fermentation of otherwise unused lignocellulosic material is an attractive alternative towards facing the worldwide challenges of energy supply and resource shortage \cite{Carroll2009}. However, further research is required before the process can be economically viable in industrial use. In particular, questions regarding the effect of the lignocellulose structure on the action of the enzymes as well as their interaction with non-digestible components remain insufficiently answered. While saccharification is intensively investigated from an experimental angle \cite{Weidener2020,Bura2009,Kucharska2018,Zoghlami2019,Pena2019,Thomas2013}, computational models are so far focused mainly on the static structure of lignocellulosic material \cite{Charlier2012,Yi2012,Ciesielski2020}, and its digestion is less studied \cite{Vetharaniam2014} \cite{Kumar2017}.\\

\noindent In this study we have built and analyzed a computational model which simulates the dynamics of the enzymatic saccharification of a cellulose microfibril surrounded by hemicellulose and lignin, which is spatially resolved at the scale of substrate monomers. The model considers both the arrangement of the polymers and effects such as lignin adhesion and the crystallinity of the substrate. It relies on a stochastic Gillespie algorithm to simulate the saccharification dynamics of the system over time. Furthermore, it offers to freely tune the enzyme cocktail composition in terms of cellulases and xylanases, and to investigate both their individual action and their synergism. Even though the model is a coarse-grained and simplified representation of the \textit{in vivo} system, it nonetheless retains essential features. It semi-quantitatively reproduces experimental data, and even suggests new explanations for their interpretation. Thereby, we demonstrate the strength of the model and its reuse potential for theory enriched experimental analyses.\\

\noindent Within the model, we keep track of all polymers and thus visualize the action of the enzyme cocktail in great detail (see Figure \ref{Heatmaps}). In particular, we can underline the current understanding of the synergism exhibited by them. Cellobiohydrolase (CBH) can only digest the ends of cellulose polymers, and therefore requires endoglucanase (EG) to generate more of these ends by cutting the initially long polymers into a large number of shorter ones. On the other hand, CBH is required to efficiently generate cellobiose. Finally, \textbeta -glucosidase (BGL) is the enzyme leading to the release of glucose. We have simulated the action and dynamics of the cellulases in detail, but have treated the action of xylanases in a coarse-grained fashion by only including a single representative xylanase enzyme (XYL). A possible expansion of the model would be to consider the diversity and complexity of both the xylanases and the hemicellulose, for instance by including different types of sugars and the corresponding xylanase sub-types.\\

\noindent We have implemented known effects of lignin into our model, i.e., its adhesive properties towards enzymes and its structural blocking function. The analysis of the influence of these effects shows an inverse linear dependence between lignin content and final saccharification yield, similarly to that found by Chen and Dixon \cite{Chen2007} and Studer et al. \cite{Studer2011} (see Figure \ref{lignin_percentage_vs_saccharification}). The model additionally suggests that the dependence may only be linear within a finite range of lignin content that is the intermediate regime of a more complex logistic decay. Upon selectively activating and deactivating the effects associated to lignin (adhesion and structural blocking), we show that when the lignin adhesion effect is active the simulated saccharification yield strongly depends on the relative amount of enzymes and lignin in the system. This result directly reflects the fact that the adhesion capacity of lignin depends on its available surface. In addition to investigating the saccharification yield dependence on the overall lignin content, Studer et al. sorted their lignocellulosic materials into two groups \cite{Studer2011}: those whose lignin S/G ratio (S: syringyl; G: guaiacyl)  was below 2.0, and those whose S/G ratio was above 2.0. They observed a steeper linear dependence of the saccharification yield on the lignin content for lower S/G ratios. This suggests that guaiacyl units have a stronger influence on the recalcitrance of the material, possibly because they have a higher affinity for binding enzymes than syringyl units do. As another future direction, we could investigate and possibly quantify this 
by diversifying the representative monolignol, which is currently the placeholder for lignin monomers in our model. The latter could be replaced with defined S- and G- units, which would exhibit differing adhesion strength.\\

\noindent We choose to define the substrate crystallinity as the inverse of the digestibility by splitting the substrate into easily digestible "amorphous" and difficult to digest "crystalline" regions. Our results show an inverse linear dependence between the cellulose crystallinity and the final saccharification yield, which is in good agreement with experiments done by Cui et al. \cite{Cui2014} and Pena et al. \cite{Pena2019} (see Figure \ref{Various_crystallinities_figure}). A further point of interest is that Cui et al. have proposed that the digestibility of crystalline cellulose does not only depend on the crystallinity index measured via x-ray diffraction (the analogue to the crystallinity fraction within our model) \cite{Cui2014}. They suggest an additional contribution by the crystal allomorph of the cellulose substrate, i.e., the shape of the microfibril. This effect of the allomorph type on the saccharification dynamics could be investigated within our model by changing the shape of the simulated microfibril.\\

\noindent We investigated the impact of the structural properties of the substrate by attempting to reproduce experimental time courses by Bura et al. \cite{Bura2009} in three different situations: a substrate without structure, a substrate with structure but without crystallinity, and a substrate with both structure and crystallinity. As shown in Figure \ref{Bura_comparison}, we were unable to find suitable parameters to reproduce the experimental data for the first two situations, but obtained good agreement for the third. This indicates that, while the substrate composition and the arrangement of the polymers are important, they are not sufficient to fully interprete the saccharification dynamics, which additionally requires to consider the crystallinity of the substrate. Our results also show that the crystallinity is reduced with increasing pre-treatment severity. Furthermore, the inclusion of crystallinity parameters for both cellulose and hemicellulose was necessary to obtain good agreement with the experimental data. This supports the semi-crystalline arrangement of xylan around cellulose microfibrils, which was proposed by Simmons et al. \cite{Simmons2016}.\\

\noindent The model captures some of the essential properties of lignocellulose saccharification that are known or hypothesized in literature. We are confident that its flexibility makes it a general platform that allows vast possibility of further development. For instance, the model could permit us to investigate different plant mutants, tissues and enzyme abundances, and kinetics.

\section*{Acknowledgements}

The authors would like to thank Prof. Dr Markus Pauly, Dr Merve Seven and Dr Vicente Ramirez Garcia for helpful discussions at the early stage of this study. The current position of E.B. is funded by the Deutsche Forschungsgemeinschaft (DFG) under Germany’s Excellence Strategy EXC 2048/1, Project ID: 390686111. The current position of A.R. is funded by the Federal Ministry of Education and Research of Germany in the framework of CornWall (Project Number 031B0193A).

\section*{Conflict of interest}

The authors have no conflict of interest to declare.

\section*{Author Contributions}

\noindent\textbf{E.B.} Conceptualization, Methodology, Software, Validation, Formal analysis, Investigation, Data Curation, Writing - Original Draft, Writing - Review \& editing, Visualization.\\

\noindent\textbf{A.R.} Conceptualization, Methodology, Validation, Resources, Writing - Original Draft, Writing - Review \& editing, Supervision, Project administration, Funding acquisition.

\end{multicols}

\newpage
\appendix
\section{Enzyme concentration}\label{Enzyme concentration}

\noindent Our aim here is to verify that when we compare experimental setups to our computational system, it yields reasonable enzyme concentrations. To do so, we consider both the geometrical properties of our \textit{in silico} substrate and the experimental setup used by Bura et al. \cite{Bura2009}. We do this by assuming that their macroscopic substrate is divided into identical sub-units, which each resemble the simulated structure in our model. Therefore, each sub-unit represents a cellulose microfibril surrounded by an outer shell of hemicellulose and lignin. Furthermore, the length of a sub-unit is equal to the length of the cellulose microfibril it contains.\\

 \noindent Since we consider our \textit{in silico} volume as unity, the enzyme concentration ($c_\textrm{enzyme}$) has the same value as the number of enzymes ($N_\textrm{enzyme}$) floating in the volume surrounding a single sub-unit ($V_\textrm{surrounding}$). Here we fix the number of enzymes to 100, and now we want to deduce the corresponding concentration by calculating the volume surrounding a single sub-unit.\\

\noindent The volume surrounding a single sub-unit is

\vspace{.5cm}

\begin{equation}
V_\textrm{surrounding} = \frac{ V_\textrm{solution} - V_\textrm{SU,tot}}{N_\textrm{SU}}\,,
\label{Volume_surrounding_eq}
\end{equation}

\vspace{.5cm}
\noindent where $V_\textrm{solution}$ is the total volume of the solution, $V_\textrm{SU, tot}$ is the volume occupied by all the sub-units, and $N_\textrm{SU}$ is their number. Within the saccharification experiments done by Bura et al., the solution volume for each sample is $V_\textrm{solution} = 50$\,ml \cite{Bura2009}. Thus, to determine $V_\textrm{surrounding}$ we require $N_\textrm{SU}$ and $V_\textrm{SU, tot}$.\\

\noindent \textbf{Number of sub-units.} Considering that all sub-units are identical, the number of sub-units within the substrate is:

\vspace{.5cm}

\begin{equation}
N_\textrm{SU} = \frac{m_\textrm{substrate}}{m_\textrm{SU}}\,,
\end{equation}

\vspace{.5cm}

\noindent where $m_\textrm{SU}$ is the mass of a single sub-unit and $m_\textrm{substrate}$ is that of the whole substrate. In the study by Bura et al., the substrate was diluted at a so-called weight-to-volume consistency of $8\%$, which corresponds to a total substrate mass of $m_\textrm{substrate} = $ 4g. \\

\noindent The mass of a single sub-unit depends on the abundance and molecular masses of its constituents, following: 

\vspace{.5cm}

\begin{equation}
m_\textrm{SU} = m_\textrm{glc} \cdot N_\textrm{glc} + m_\textrm{xyl} \cdot N_\textrm{xyl} + m_\textrm{lign} \cdot N_\textrm{lign}\,.
\label{microfibril_mass_eq}
\end{equation}

\vspace{.5cm}

\noindent The numbers of monomers ($N_\textrm{glc}$, $N_\textrm{xyl}$, and $N_\textrm{lign}$) are calculated from the sub-unit length and the percentages of cellulose, hemicellulose and lignin. We use the percentages found by Bura et al. for "medium pre-treatment severity" (see Table \ref{Param_table}). The molecular masses for the three monomers ($m_\textrm{glc}$, $ m_\textrm{xyl}$, and $m_\textrm{lign}$) are shown in Table \ref{Molecular_weights_table}.\\

\noindent \textbf{Volume of sub-units.} The volume of all sub-units together is:

\vspace{.5cm}

\begin{equation}
V_\textrm{SU,tot} = V_\textrm{SU} \cdot N_\textrm{SU}\,,
\label{fibril_volume_eq}
\end{equation}

\vspace{.5cm}

\noindent where $V_\textrm{SU}$ is the volume occupied by a single sub-unit, and is calculated following:

\vspace{.5cm}

\begin{equation}
V_\textrm{SU} = d_\textrm{SU}^2 \cdot (N_\textrm{bonds}+1) \cdot d_\textrm{glc}\,.
\label{fibril_volume_eq}
\end{equation}

\vspace{.5cm}

\noindent $d_\textrm{glc}$ is the diameter of a glucose molecule, assumed to be equal to the distance between bonds, $d_\textrm{SU}$ is the diameter of a sub-unit, and $N_\textrm{bonds}$ is its length, counted in bonds. First, we assume that $d_\textrm{glc}$ is roughly 1 nm. Second, Thomas et al. estimate that the side length of a rectangular cellulose microfibril of 24 polymers is roughly 3 nm \cite{Thomas2013}. In order to include the outer layers of hemicellulose and lignin, we set $d_\textrm{SU} = 5$\,nm. Third, the bond number $N_\textrm{bonds}$ is set to 100.\\

\noindent Consequently, we can calculate the volume surrounding a single sub-unit, $V_\textrm{surrounding} \approx 1.4 \cdot 10^{-20}\,\textrm{m}^3$, and the concentration of enzymes is

\vspace{.5cm}

\begin{equation}
c_\textrm{enzyme} = \frac{N_\textrm{enzyme}}{V_\textrm{surrounding}} \approx 10\,\textrm{\textmu M.}
\label{Enzyme_concentration_assumption_eq}
\end{equation}

\vspace{.5cm}

\noindent This concentration is of a magnitude that is within the range of some known \textit{in vivo} enzyme concentrations \cite{Albe1990}, and we therefore consider it reasonable.

\begin{table}[ht]
	\centering
	
	\begin{tabular}{|l|l|l|l|}
		\hline
		\textbf{Molecule} & Glucose & Xylose & Monolignol \\      \hline
		\textbf{Molecular mass ($\frac{\textrm{kg}}{\textrm{mol}}$)} & 0.180 & 0.150 & 0.180\\      \hline
	\end{tabular}
	\captionof{table}{
		Molecular masses of the constituents of the lignocellulose sub-units, obtained from literature and rounded to three digits. The value for a lignin monomer, called a monolignol, was taken as the mean value of the molecular masses of the three main monolignols (coumaryl alcohol, coniferyl alcohol and sinapyl alcohol).
	}
	\label{Molecular_weights_table}
\end{table}

\end{document}